\documentclass[showpacs,prb,superscriptaddress,twocolumn]{revtex4}

\usepackage{amsfonts}
\usepackage{graphicx}
\usepackage{amsmath}
\usepackage{amssymb}
\usepackage{latexsym}
\usepackage{psfrag}

\begin{document}

\title{Dirac Point Resonances due to Atoms and Molecules Adsorbed  
on Graphene
and Transport Gaps and Conductance Quantization in Graphene Nanoribbons 
with Covalently 
Bonded Adsorbates}
\author{Siarhei Ihnatsenka}
\affiliation{Department of Physics, Simon Fraser University, Burnaby, 
British Columbia, Canada V5A 1S6}
\author{George Kirczenow}
\altaffiliation{Canadian Institute for Advanced Research, 
Nanoelectronics Program.}
\affiliation{Department of Physics, Simon Fraser University, 
Burnaby, British Columbia, Canada V5A 1S6}

\begin{abstract}
We present a tight binding theory of the Dirac point resonances due to adsorbed
atoms and molecules on an infinite 2D graphene sheet based on the standard tight
binding model of the graphene $\pi$-band electronic structure and the extended
H\"{u}ckel model of the adsorbate and nearby graphene carbon atoms. The relaxed
atomic geometries of the adsorbates and graphene are calculated using density
functional theory. Our model includes the effects of the local rehybridization
of the graphene from the $sp^2$ to $sp^3$ electronic structure that occurs when
adsorbed atoms or molecules bond covalently to the graphene. Unlike in previous
tight-binding models of Dirac point resonances, adsorbed species with multiple
extended molecular orbitals and bonding to more than one graphene carbon atom
are treated. More accurate and more general analytic expressions for the Green's
function matrix elements that enter the $T$-matrix theory of Dirac point
resonances than have been available previously are obtained. We study H, F, OH
and O adsorbates on graphene and for each we find a strong scattering resonance
(two resonances for O) near the Dirac point of graphene, by far the strongest
and closest to the Dirac point being the resonance for H. We extract a minimal
set of tight binding parameters that can be used to model resonant electron
scattering and electron transport in graphene and graphene nanostructures with
adsorbed H, F, OH and O accurately and efficiently. We also compare our results
for the properties of Dirac point resonances due to adsorbates on graphene with
those obtained by others using density functional theory-based electronic
structure calculations, and discuss their relative merits. We then present
calculations of electronic quantum transport in graphene nanoribbons with these
adsorbed species.  Our transport calculations capture the physics of the
scattering resonances that are induced in the graphene ribbons near the Dirac
point by the presence of the adsorbates. We find the Dirac point resonances to
play a dominant role in quantum transport in ribbons with adsorbates: Even at
low adsorbate concentrations the conductance of the ribbon is strongly
suppressed and a transport gap develops for electron Fermi energies near the
resonance. The transport gap is centered very near the Dirac point energy of for
H, below it for F and OH and above it for O. We find conduction in ribbons with
adsorbed H atoms to be very similar to that in ribbons with equal concentrations
of carbon atom vacancies. We predict ribbons with adsorbed H, F, OH and O, under
appropriate conditions, to exhibit quantized conductance steps of equal height
similar to those that have been observed by Lin {\em et al.} [Phys. Rev. B
\textbf{78}, 161409(R) (2008)] at moderately low temperatures, even for ribbons
with conductances a few orders of magnitude smaller than $2e^2/h$.

\end{abstract}


\maketitle

\section{Introduction}
\label{Introduction}

In recent years graphene nanoribbons have been the subject of increasing
experimental\cite{Han07,Chen07, Lin08,Li08sonif,Wang08, Molitor09, Stampfer09,
Koskinen09, Jiao09, Todd09, Kosynkin09,Han2010,
Gallagher2010,Oostinga2010,Jiao2010,Cai2010} and
theoretical\cite{Dresselhaus96,Wakabayash01,Hikihara03,Lee05,Brey06,Son06,
Peres06,Barone06,Areshkin07,Gunlyckea07,
Sols07,Pisani07,Yang07,Gunlyckeb07,Onipko08,Igor08,Evaldsson08,theory,
Yamamoto08,Jiang08,Dutta08a,Mucciolo09,corrugated09,disorder09,Lopez09,LaMagna09,
San-Jose09, Ran09,Lin09,Dutta09,Jung09,Ramasubramaniam2010,Biel09} interest. Ideal
ribbons that are uniform in width and free of defects, adsorbates and other
disorder should transmit electrons ballistically (i.e. without scattering) and
consequently, as is the case for other quasi-one-dimensional (1D) ballistic
nanostructures,\cite{Kirczenow} their low temperature conductances are expected
to be quantized in integer multiples of
$2e^2/h$.\cite{Wakabayash01,Peres06,Areshkin07,Gunlyckea07,Onipko08, Igor08,
Evaldsson08, theory,
Yamamoto08,Mucciolo09,corrugated09,disorder09,Lopez09,LaMagna09} The ribbons
that have been realized experimentally to date have been far from ballistic;
consistent with theoretical work on strongly disordered ribbons\cite{Areshkin07,
Gunlyckea07, Sols07,Evaldsson08, Mucciolo09,
disorder09,Lopez09,LaMagna09,disorder09}  their measured conductances when a few
transverse subbands are populated with electrons have been much smaller than the
conductance quantum $2e^2/h$. Thus the  recent  experimental observation of
quantized conductance steps in graphene nanoribbons by Lin et al.\cite{Lin08} at
moderately low temperatures was surprising and especially so in view of the fact
that the observed conductance step heights were two orders of magnitude smaller
than $2e^2/h$.  In a previous paper,\cite{disorder09} we explained this puzzling
phenomenon as arising from enhanced electron backscattering near subband edge
energies due to the presence of defects. This explanation is consistent with the
conclusion drawn by Lin et al.\cite{Lin08} that the conductance steps that they
observed were evidence of subband formation in their nanoribbon samples. In the
models that we studied\cite{disorder09} carbon atom vacancies in the interior of
the ribbon played a crucial role: They were responsible for the formation of
{\em equally spaced} conductance steps in a range of temperatures $T$ high
enough to suppress universal conductance fluctuations but for which  $k_B T$ is
smaller than the subband spacing, in agreement with the experiment.\cite{Lin08}
Here $k_B $ is Boltzmann's constant. However, whether such vacancies were
actually present in the experimental samples at the required concentrations was
not determined in the experimental work of Lin et al.\cite{Lin08} and it is
widely believed based on STM\cite{STM} and TEM\cite{TEM} measurements that
graphene samples can be free from carbon atom vacancies over large areas. On the
other hand it is reasonable to expect atomic and molecular species to be
adsorbed on graphene ribbons prepared using presently available fabrication
techniques.\cite{Stojkovic03,Leenaerts2008,Leenaerts2009,Leenaerts2009a,Ao2010,
Ao2010a,Lopez09} Thus it is of interest to explore the possible role that such
adsorbates may play in conductance quantization of the kind reported by  Lin et
al.\cite{Lin08}

A remarkable property of pristine graphene is that the electron dispersion near
the Fermi energy is linear and forms Dirac-like cones in $k$-space centered on
two points $K$ and $K'$ in the Brillouin zone.\cite{review} The energy at which
the density of states  of graphene vanishes is known as the Dirac point. It has
been suggested  that impurities that strongly perturb the graphene should give
rise to resonant states in the vicinity of the Dirac point and that these
resonant states result in strong scattering of electrons in the
graphene.\cite{Skrypnyk06,Pereira06, Wehling07,  Robinson08, Pereira08,
Wehling09X, Wehling09, Wehling10,Skrypnyk07, Basko08, Pershoguba09, Skrypnyk10} 
Adsorbed atoms and molecules that are covalently bonded to graphene perturb the
graphene strongly and thus may be expected to give rise to such ``Dirac point
resonances". However, how adsorbate-induced Dirac point resonances affect
electron transport in graphene {\em nanoribbons} is a topic that is yet to be
explored theoretically or experimentally.

In this paper we generalize the previously proposed analytic theories of the
impurity-induced Dirac point resonances\cite{Skrypnyk06, Pereira06, Wehling07, 
Robinson08, Pereira08, Wehling09X, Wehling09, Wehling10, Skrypnyk07, Basko08,
Pershoguba09, Skrypnyk10} to the case of adsorbates on graphene whose electronic
structure is described within the extended H\"{u}ckel  model of quantum
chemistry.\cite{yaehmop} The extended H\"{u}ckel model\cite{Kirczenow}  is a
semi-empirical tight binding scheme that provides a simple but reasonably
realistic description of the electronic structures of many molecules. It has
been used successfully in explaining and predicting experimental transport
properties of a variety of molecular systems\cite{Kirczenow} including
conduction in molecular wires bridging metal contacts\cite{Datta1997,
EmberlyKirczenow01,  EmberlyKirczenow01PRB, Kushmerick02, Cardamone08,
Cardamone10}  and molecular arrays on silicon\cite{PivaWolkowKirczenow05,
PivaWolkowKirczenow08, PivaWolkowKirczenow09} and the
electroluminescence,\cite{Buker08} current-voltage characteristics\cite{Buker08}
and STM images\cite{Buker05} of molecules on complex substrates. Thus it offers
a natural way to extend the standard tight binding model of pristine
graphene\cite{review}  to the case of graphene with adsorbates.
 
Here we develop a tight binding model of graphene with adsorbates based on
extended H\"{u}ckel theory and use it to carry out quantum transport
calculations for graphene nanoribbons with adsorbates. An important advantage of
our approach is that it includes {\em multiple} atomic or molecular orbitals of
the adsorbed species as well as the relevant non-$\pi$ graphene orbitals
explicitly in the tight binding model used to calculate the effect of Dirac
point resonances on transport whereas other tight-binding transport calculations
\cite{Robinson08, Wehling10, Skrypnyk10} (which have been for {\em 2D graphene}
with adsorbates or impurities) have been restricted to much simpler {\em
single}-orbital models of Dirac point resonances.

As specific examples of adsorbates we consider H, F and O atoms and OH groups,
species that may have been present in the experimental samples of Lin et
al,\cite{Lin08} that were made by oxygen plasma reactive ion etching using a
hydrogen silsesquioxane etch mask that was later removed in a hydrofluoric acid
solution.

We estimate the relaxed geometries of these adsorbates and carbon atoms to which
they bond using {\em ab initio} density functional calculations.\cite{gaussian}
It is energetically favorable for the graphene carbon atoms to which adsorbed
species bond to move out of the graphene plane towards the adsorbate by
fractions of an Angstrom and partial rehybridization of the carbon atom from
$sp^2$ to $sp^3$ bonding then occurs.\cite{Stojkovic03,Wehling09,
Igami01,Lopez09} Therefore in our treatment of the Dirac point resonances we
include the atomic valence orbitals of the carbon atoms that are involved in the
 $sp^3$ bonding {\em in addition} to the atomic valence orbitals of the adsorbed
species and the  $2p_z$ orbitals of the graphene carbon atoms that are included
\cite{review} in the standard tight binding model of graphene. Our theory also
applies to species that bond simultaneously to more than one graphene carbon
atom as is the case for an adsorbed O atom.

We develop more accurate and more general analytic expressions than have been
available to date\cite{Skrypnyk06,Pereira06, Wehling07,  Robinson08, Pereira08,
Wehling09X, Wehling09, Wehling10,Skrypnyk07, Basko08, Pershoguba09, Skrypnyk10}
for the matrix elements of the Green's function of pristine graphene that enter
the theory of the Dirac point resonances and check their accuracy by numerical
calculations. We then calculate the matrix elements of the $T$-matrix that
describes the scattering of graphene electrons by the adsorbate as a function of
energy and thus determine the energies at which the Dirac point resonances of
the various adsorbates occur as well as the resonance energy profiles.

For each of the H, F and OH adsorbed species we find a strong resonance located
near the Dirac point. For the adsorbed O atom we find the $T$-matrix to exhibit
a more complex energy profile with a pair of overlapping resonances of different
widths near the Dirac point. For each of these adsorbed species we also develop
a minimal set of tight binding parameters that yield an accurate description of
its Dirac point resonance(s). These parameter sets are used in our transport
calculations on graphene nanoribbons that we report here and are expected also
to be useful in other theoretical work such as studies of the tunneling spectra
of adsorbates on graphene that may be observed in scanning tunneling
spectroscopy experiments.

We show that Dirac point resonances due to adsorbates have a strong signature in
the transport characteristics of graphene nanoribbons that depends strongly on
the adsorbed species even at low adsorbate concentrations. We investigate the
possible influence of adsorbates on the quantized conductances that have been
observed experimentally\cite{Lin08} in graphene nanoribbons at moderately low
temperatures, and that we have studied theoretically previously \cite
{disorder09} in models of graphene nanoribbons with vacancies. The results
obtained here regarding adsorbate-induced Dirac point resonances in graphene
also provide a physical interpretation of key features of the calculated
nanoribbon conductances that we shall present. The transport calculations that
we present yield the following salient results::

(i) The adsorption of each of the species that we study on a graphene ribbon
results in very strong electron scattering especially at energies in the
vicinity of the Dirac point scattering resonance associated with the respective
adsorbate. This in turn leads to strong suppression of the ribbon conductance
and a transport gap opening up in a range of electron Fermi energies near the
energy of the Dirac point resonance.

(ii) The transport gaps occur for electron Fermi energies around the Dirac point
for H, below the Dirac point for F and OH and above the Dirac point for O
adsorbates. The case of O differs qualitatively from those of F and OH because O
binds to two carbon atoms belonging to different graphene sublattices while the
F and OH  bind to a single carbon atom.

(iii) The conductance characteristics as a function of the electron Fermi energy
for ribbons with adsorbed H atoms are very similar both qualitatively and
quantitatively to those of ribbons with the same concentration of carbon atom
vacancies.

(iv) We predict that ribbons with each of these adsorbed species should under
appropriate conditions exhibit equally spaced conductance steps at moderately
low temperatures even for adsorbate concentrations for which the conductance is
much smaller than $2e^2/h$, consistent with experiment.\cite{Lin08}

The remainder of this article is organized as follows. In Sections  \ref{tbH} -
\ref{DFTvsEH} we present a theory of the {\em infinite 2D} graphene sheet with
adsorbed atoms or molecules. Then in Sections \ref{model} and \ref{results} we
apply the results obtained in the preceding sections to the problem of transport
in graphene {\em nanoribbons}. We formulate our tight binding model of the
electronic structure of adsorbed atoms and molecules on graphene in Section
\ref{tbH}. In Section \ref{effective} we show how the tight-binding Hamiltonians
can be transformed into effective graphene Hamiltonians that include the effect
of the adsorbate by generalizing previous theoretical work to the case of
adsorbate species with multiple molecular orbitals and bonding to more
than one graphene carbon atom. In Section \ref{Tmatrix} we briefly discuss the
$T$-matrix theory used to study Dirac point resonances in graphene analytically.
In Section \ref{analyticG} we derive an exact relation between the graphene
Green's function matrix elements that enter the $T$-matrix theory. We use this
relation to obtain accurate analytic expressions for these matrix elements and
check their accuracy by exact numerical calculations. In Section \ref{Dirac} we
present our results for the Dirac point resonances of H, F and O atoms and OH
groups on graphene as well as a minimal tight binding model that accurately
reproduces those results and is used in the transport calculations that we 
present in Section \ref{results}. 
In Section \ref{Dirac} we also examine quantitatively the effect of the
adsorbate-induced  $sp^3$ rehybridization of the graphene on the Dirac point
resonances and find it to be important, and especially so for adsorbed H atoms.
In Section \ref{DFTvsEH} we compare our results for the properties of Dirac
point resonances due to adsorbates on graphene with those obtained by others
using density functional theory-based electronic structure calculations. In
Section \ref{model} we discuss the model and methodology used in our
calculations of transport in graphene {\em nanoribbons}. In Section
\ref{results} we present: In Section \ref{H_Adatoms} we discuss
ribbons with H adatoms including the effects of Dirac point resonances and
graphene rehybridization on transport and on the electronic density of states.
Ribbons with F, OH and O adsorbates are then considered in Sections
\ref{F_Adatoms}, \ref{OH_Ads} and \ref{O_Ads}. We comment on conductance
asymmetries of ribbons relative to the Dirac point that result from the presence
of adsorbates in Section \ref{asym}. In Section \ref {Quantization} we focus on
the topic of conductance quantization in ribbons with adsorbates. We discuss
renormalization of the energies of the ribbon subbands and Dirac point due to
the presence of adsorbates in Section \ref{renorm} and the dependence of the
transport gaps on the adsorbate concentration in Section \ref{gapsvsconc}. In
Section  \ref{Discussion} we summarize our main findings and comment briefly on
potential relevant experiments.

\section{The Tight Binding Hamiltonian}
\label{tbH}

Our starting point is the simplest tight-binding model of pristine 
graphene\cite{Wallace47} embodied in the Hamiltonian

\begin{equation}
 H_0= - \sum_{\left\langle i,j\right\rangle }t\left( a_{i}^{\dag }a_{j}+
 \mathrm{h.c.} \right) 
 \label{pristine}
\end{equation}%
Here $-t$ is the Hamiltonian matrix element between nearest-neighbor $2p_z$
carbon orbitals of the graphene lattice and $a_{i}^{\dag }$ is the creation
operator for an electron in $2p_z$ carbon orbital $i$. This Hamiltonian with
$t=2.7$ eV is known to describe the $\pi$ band dispersion of graphene well at
energies around the Dirac point,\cite{Reich02, review} i.e., in the energy range
of interest in the present work. We extend this tight binding model to include
the adsorbate and its coupling to the graphene carbon atoms in the following
way.

\begin{figure}[t]
\includegraphics[scale=0.35]{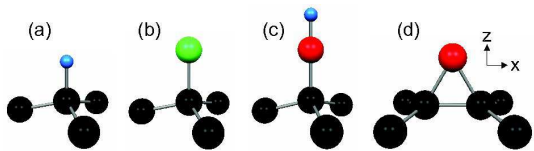}
\caption{(color online) Relaxed geometries of adsorbates on graphene. C, H, F,
and O atoms are black, blue, green and red respectively.\cite{Macmolplt}
(a)Adsorbed hydorgen atom. H atom and C atom to which H binds are 1.47 and 0.35
{\AA} above graphene plane. (b) Adsorbed fluorine.  F atom and C atom to which F
binds are 1.83 and 0.36 {\AA} above graphene plane. (c) Adsorbed hydroxyl group.
H atom, O atom and C atom to which O binds are 2.78, 1.83 and 0.41 {\AA} above
graphene plane. (d) Adsorbed oxygen.  O atom and C atoms to which O binds are
1.51 and 0.27 {\AA} above graphene plane. The C atoms to which the O binds are
separated by 1.47 {\AA}. The $x$-axis is parallel to the line joining the C
atoms to which the O binds.}
\label{geom}
\end{figure}

We performed {\em ab initio} geometry relaxations based on density functional
theory for the adsorbed species on the honeycomb graphene lattice using the
Gaussian 09 software package.\cite{gaussian} The relaxed geometries calculated
in this way are expected to be accurate since density functional theory has been
well optimized for carrying out accurate ground state total energy calculations
on which these relaxations are based.\cite{Kirczenow} The structures studied
were graphene disks of several tens of carbon atoms passivated at the edges with
hydrogen, the adsorbed species being bonded to the graphene near the center of
the disk.  The atoms of the adsorbed species and the carbon atoms with which
they bond were allowed to relax freely, the other carbon atoms being held fixed
in the standard hexagonal graphene geometry with the C-C distance of 1.42 \AA.
The relaxed structures obtained in this way are shown in Fig. \ref{geom}.\cite{Macmolplt} The
tight binding model Hamiltonian Eq. (\ref{pristine}) was extended to include the
atomic valence orbitals of the adsorbate and their coupling to the valence
orbitals of the graphene carbon atoms by calculating the relevant matrix
elements within the extended H\"{u}ckel  model. \cite{Kirczenow}

Extended H\"{u}ckel theory is formulated in terms of small basis sets of 
Slater-type atomic orbitals $\left\{\left|\phi_i\right\rangle\right\}$, their
overlaps $S_{ij} = \left\langle \phi_i | \phi_j \right\rangle$, and a 
Hamiltonian 
matrix $\mathcal{H}_{ij} = \left\langle \phi_i \left|\mathcal{H} \right| 
\phi_j \right\rangle$. The diagonal Hamiltonian elements $\mathcal{H}_{ii}= 
\mathcal{E}_i$ are chosen to be the experimentally determined atomic orbital 
ionization energies $\mathcal{E}_i$. In the present work the nondiagonal 
elements 
are approximated as in Ref.  \onlinecite{yaehmop} by 
$\mathcal{H}_{ij} = (1.75 + 
\Delta_{ij}^2 - 0.75  \Delta_{ij}^4)  S_{ij} \left( \mathcal{E}_i + 
\mathcal{E}_j \right)/2$, where $\Delta_{ij}=( \mathcal{E}_i - 
\mathcal{E}_j)/( \mathcal{E}_i + \mathcal{E}_j)$, a form chosen\cite{yaehmop} 
for consistency with experimental molecular electronic structure data. In the 
standard tight binding Hamiltonian (\ref{pristine}) of pristine graphene, the 
energy scale  is chosen so that the carbon $2p_z$ orbital energy is zero 
whereas in extended H\"{u}ckel theory\cite{yaehmop} the carbon $2p_z$ orbital 
energy is the ionization energy $ \mathcal{E}_{\mathrm{C}_{p_z}} = -11.4$ eV.
Accordingly, for consistency, in our extended  H\"{u}ckel Hamiltonian matrix we
make the replacement $\mathcal{H}_{ii} \rightarrow \mathcal{H}_{ii}-
\mathcal{E}_{\mathrm{C}_{p_z}}$. Because the extended H\"{u}ckel  basis states
on different atoms are not in general mutually orthogonal the non-diagonal
extended  H\"{u}ckel Hamiltonian matrix elements are then also adjusted 
according to
\begin{equation}
\mathcal{H}_{ij}  \rightarrow  \mathcal{H}_{ij} - S_{ij} 
\mathcal{E}_{\mathrm{C}_{p_z}}
\label{eq:shift}
\end{equation}%
as is discussed in Ref.\onlinecite{shift}.  

Let 
$\mathcal{H}_{ij}^\mathrm{R}$ and  $S_{ij}^\mathrm{R}$ be the  extended
H\"{u}ckel Hamiltonian and overlap matrices defined in this way but restricted
to the Hilbert subspace R spanned by the valence orbitals of the adsorbate and
valence orbitals of the graphene {\em other than} the 2$p_z$ graphene orbitals
that in the present model are already described by $H_0$. In the numerical
results presented in this paper in addition to the adsorbate valence orbitals we
include in R the 2$s$, $2p_x$ and $2p_y$ orbitals (see Fig.\ref{geom}) of the
carbon atom(s) to which the adsorbed atom or molecule bonds and of its three
nearest carbon atom neighbors. We calculate $\mathcal{H}_{ij}^\mathrm{R}$ and 
$S_{ij}^\mathrm{R}$ for the relaxed geometries shown in Fig. \ref{geom}. We then
solve the extended H\"{u}ckel Schr\"{o}dinger equation
\begin{equation}
\mathcal{H}^\mathrm{R} \psi_\alpha = \epsilon_\alpha S^\mathrm{R} \psi_\alpha
\label{restricted}
\end{equation}%
(for a single adsorbed H, F or O atom or OH group) numerically for its
eigenstates  $\psi_\alpha$ and energy eigenvalues $\epsilon_\alpha$. The 
eigenstates  $\psi_\alpha$ obtained in this way are mutually orthogonal. They
should be regarded as {\em extended molecular orbitals} (EMOs) of the adsorbate.
They are linear combinations of the atomic valence orbitals of the adsorbate and
{\em some} of the atomic valence orbitals of the graphene (as is detailed above
Eq. \ref {restricted}) but do {\em not} include any graphene $2p_z$ atomic
orbitals.  The EMOs will play a central role in the theory that follows.

The EMOs $\psi_\alpha$ together with the $2p_z$ orbitals of the graphene carbon
atoms form the basis set for our tight binding Hamiltonian $H$ of the
graphene-adsorbate system that we write in the form
\begin{equation}
 H= H_0 + \sum_{\alpha}\epsilon_{\alpha}d_{\alpha}^{\dag }d_{\alpha} + 
 \sum_{\alpha, j} \gamma_{\alpha j} \left( d_{\alpha}^{\dag }a_{j}+h.c. 
 \right) ,
 \label{eq:hamiltonian}
\end{equation}%
Here $a_{j} $ is the destruction operator for an electron in the $2p_z$ orbital
$\phi_j$ of carbon atom $j$. $d_{\alpha}^{\dag }$ is the creation operator for
an electron in EMO $\psi_\alpha$ that is an eigenstate of Eq. (\ref{restricted})
and $\epsilon_{\alpha}$ is the corresponding energy eigenvalue. 
$ \gamma_{\alpha j} = \langle \psi_\alpha | \mathcal{H}  | \phi_j \rangle$ 
the matrix element of
the extended H\"{u}ckel Hamiltonian between the $2p_z$ orbital $\phi_j$ of
carbon atom $j$ and  EMO  $\psi_\alpha$. For simplicity, in this paper we
include in the sum over $j$ in the last summation on the RHS of Eq. 
(\ref{eq:hamiltonian}) only the $2p_z$ orbital of the carbon atom that is
closest to the adsorbed moiety (or in the case of the adsorbed O the closest two
carbon atoms), and we also neglect any changes in $t$ in Eq. (\ref{pristine})
that occur due to the change in the graphene geometry induced by the adsorbate.
The latter latter effect is however taken into account in the numerical
nanoribbon  transport calculations that we report in Section \ref{results}.

We note that with the above definitions, the couplings between all of the
valence orbitals ($2s, 2p_x, 2p_y$ and $2p_z$) of the carbon atom to which the
adsorbate bonds and all of the $2s, 2p_x, 2p_y$ and $2p_z$ valence orbitals of
that carbon atom's nearest carbon atom neighbors are included in the present
model that is summarized by Eq.  (\ref{eq:hamiltonian}). In this way we include
in our calculation all of the carbon atom valence orbitals that participate in
the adsorbate induced $sp^3$ bonding.

Because the basis set used in extended H\"{u}ckel theory is non-orthogonal the
overlap $ \sigma_{\alpha j} = \langle \psi_\alpha  | \phi_j \rangle$ between the
$2p_z$ orbital $\phi_j$ of carbon atom $j$ and  EMO  $\psi_\alpha$ may be
non-zero. This overlap is neglected in Eq.(\ref {eq:hamiltonian}). It has been
shown\cite{orthog1,orthog2} that transport problems formulated in a
nonorthogonal basis can be solved by transforming to an alternate Hilbert space
in which the basis is orthogonal but the effective Hamiltonian matrix elements
become energy-dependent. This transformation is the foundation of the standard
methods used today to treat basis set non-orthogonality throughout the molecular
electronics transport literature. For the present system, the transformation is
accomplished by replacing $ \gamma_{\alpha j}$ in Eq. (\ref{eq:hamiltonian}) by
$ \gamma_{\alpha j}-\epsilon \sigma_{\alpha j}$. Here $\epsilon$ is the electron
energy at which the Landauer electron transmission probability through the
system is calculated.  This correction is included in the numerical results that
we present in this paper for the adsorbate induced Dirac point resonances,
although in the interests of clarity it will not appear explicitly in the
formulae that we present in the remainder of this article.

\section{Effective Hamiltonians}
\label{effective}

 In the simplest possible model of an adsorbate represented by just one atomic
 orbital $\alpha$ that couples only to the $2p_z$ orbital of only one carbon
 atom $j$ of the graphene  the tight binding Hamiltonian  of the graphene and
 adsorbate is $H_1=  H_0 + \epsilon_{\alpha}d_{\alpha}^{\dag }d_{\alpha} + 
 \gamma_{\alpha j} \left( d_{\alpha}^{\dag }a_{j}+\mathrm{h.c.} \right) $ where
 the notation is as in Eq. \ref{eq:hamiltonian}. The eigenstate  $|\Psi\rangle$
 of $H_1$ with energy eigenvalue $\epsilon$ can then be written as
 $|\Psi\rangle= |\Psi_g\rangle +|\Psi_a\rangle$ where  $|\Psi_g\rangle$ and
 $|\Psi_a\rangle$ are the projections of  $|\Psi\rangle$ onto the space spanned
 by the $2p_z$ orbitals of graphene and onto the orbital of adsorbed atom,
 respectively. With these definitions, it has been shown\cite{Robinson08} that
 $|\Psi_g\rangle$ is an exact eigenstate of an effective Hamiltonian
 $H_\mathrm{eff}=  H_0 + V_{j}a_{j}^{\dag }a_{j}$ with the same energy
 eigenvalue $\epsilon$ as $|\Psi\rangle$. Here  $V_{j} = \gamma_{\alpha
 j}^2/{(\epsilon-\epsilon_\alpha)}$. Thus for the purpose of calculating the
 transport coefficients of graphene with such an adsorbed atom within Landauer
 theory, it is sufficient to replace the Hamiltonian $H_1$ with
 $H_\mathrm{eff}$, i.e, the Hamiltonian of graphene without the adsorbed atom
 but with an energy dependent potential $ \gamma_{\alpha
 j}^2/{(\epsilon-\epsilon_\alpha)}$ on carbon atom $j$ of the graphene sheet.
 
 We note that a general theory of systems with  one or more discrete states
 coupled to a continuum of states 
 was presented by Fano in 1961.\cite{Fano61} 
 Graphene with an adsorbed atom or molecule is such a system.
 Fano's analysis 
 of such systems\cite{Fano61} starts in the same way as the analysis in 
 Ref. \onlinecite{Robinson08} that we have outlined above by projecting the
 eigenstates of the system onto the continuum and discrete state manifolds. 
However, unlike in Ref. \onlinecite{Robinson08},
Fano did not reformulate the problem in terms of an effective 
Hamiltonian  $H_\mathrm{eff}$ that acts on the continuum subspace 
only.\cite{Fano61} Subsequently, effective Hamiltonians have been employed to
study bound states coupled to continua but those effective Hamiltonians
have been 
non-Hermitian operators obtained
by {\em eliminating the continuum subspace} from the theory,\cite{Bulgakov2009} 
unlike the Hermitian 
effective Hamiltonians obtained in Ref. \onlinecite{Robinson08} (and in the 
theory presented below)
by {\em eliminating the
discrete state subspace}.

In the present work we need to include more than one extended molecular orbital
$\psi_\alpha$ per adsorbed moiety in the tight binding Hamiltonian given by Eq.
(\ref{eq:hamiltonian}) for the adsorbed H, F or O atom or OH group. This is the
case even for H (which has only one valence orbital in extended H\"{u}ckel
theory) because we include several graphene atomic orbitals in the subspace R in
which we calculate the EMOs $\psi_\alpha$ for the H adsorbate as is discussed in
Section \ref{tbH}.

 We find that the argument presented in Ref.  \onlinecite{Robinson08} that leads
 to the effective Hamiltonian $H_\mathrm{eff}$ discussed above can be
 generalized in a direct way to adsorbates for which more than one effective
 molecular orbital and/or bonding of the adsorbed moiety to more than one carbon
 atom (as in the case of adsorbed O) must be considered.

 For adsorbates with more than one extended  molecular orbital that bond
 strongly to a single carbon atom, we find that  the effective Hamiltonian still
 has the form $H_\mathrm{eff}=  H_0 + V_{j}a_{j}^{\dag }a_{j}$ but the effective
 potential  $V_{j}$ becomes $V_{j} = \sum_{ \alpha } |\gamma_{ \alpha j}
 |^2/{(\epsilon-\epsilon_ \alpha )}$ where the sum is over the extended
 molecular orbitals $\alpha$ of the adsorbed moiety  that bonds to carbon atom
 $j$ of the graphene.

 For a single adsorbed O atom that bonds to two neighboring graphene carbon
 atoms 1 and 2 the effective Hamiltonian is $H_\mathrm{eff}=  H_0 +
 V_{11}a_{1}^{\dag }a_{1}+ V_{22}a_{2}^{\dag }a_{2}+ V_{12}a_{1}^{\dag }a_{2}+
 V_{21}a_{2}^{\dag }a_{1}$ where $V_{nm} = \sum_{ \alpha } \gamma^{~}_{  \alpha 
 n} \gamma^{*}_{ \alpha  m}/{(\epsilon-\epsilon_  \alpha  )}$ and the summation
 is over the extended molecular orbitals $\alpha$ of the O adsorbate. For the
 adsorbed O atom we consider 22 EMOs in this paper. They are linear combinations
 of the O 2$s$, 2$p_x$, 2$p_y$ and 2$p_z$ valence orbitals and the 2$s$, 2$p_x$
 and 2$p_y$ valence orbitals of each of the six carbon atoms shown in Fig.
 \ref{geom}(d). The effect of the adsorbed O atom on the Hamiltonian eigenstates
 projected onto the graphene $\pi$ band subspace is equivalent to the combined
 effect of energy dependent potentials applied to the 2$p_z$ valence orbitals of
 the two carbon atoms to which the O atom bonds and an energy dependent change
 in the Hamiltonian matrix elements between those 2$p_z$ carbon orbitals.

The preceding results for the effective Hamiltonians $H_\mathrm{eff}$ apply
equally to adsorbates on graphene nanoribbons or on 2D graphene. We now use them
to develop a better understanding of how H, F, OH and O adsorbates resonantly
scatter electrons by extending the general $T$-matrix approach considered
previously in Refs \onlinecite{Wehling09X, Wehling07, Skrypnyk06,
Robinson08,Skrypnyk07, Basko08, Pershoguba09, Skrypnyk10}. We consider here for
simplicity the case of an isolated single  H, F, OH or O atom or molecule on
infinite 2D graphene. In Section \ref{results} we will relate our findings to
the results of our numerical transport calculations for graphene nanoribbons.

\section{$T$-Matrix Formalism}
\label{Tmatrix}

The $T$-matrices that we consider are defined in the standard way by
\begin{equation}
 G = G^0+G^0 T G^0 
 \label{eq:G}
\end{equation}
where $G = (\epsilon +i\eta -H_\mathrm{eff})^{-1}$ is the full Green's function
based on the effective Hamiltonians $H_\mathrm{eff}$ discussed above for a
single adsorbed atom or molecule, $G^0= (\epsilon +i\eta -H_0)^{-1}$ is the
unperturbed Green's function for $\pi$ band electrons in clean graphene and $T$
characterizes the scattering strength due to the adsorbate. $T$ can be written
in the standard form
\begin{equation}
 T = \mathcal{V} + \mathcal{V}G^0\mathcal{V} + 
 \mathcal{V}G^0 \mathcal{V} G^0 \mathcal{V} + ... 
 \label{eq:Tseries}
\end{equation}
where $\mathcal{V} = a_{j}^{\dag }a_{j}\sum_{ \alpha } |\gamma_{ \alpha j}
|^2/{(\epsilon-\epsilon_ \alpha )}$ for a H, F, OH atom or molecule with EMOs
$\alpha$ bound to carbon atom \textit{j}. For an O atom with EMOs $\alpha$ bound to two
neighboring C atoms 1 and 2, $\mathcal{V} = V_{11}a_{1}^{\dag }a_{1}+
V_{22}a_{2}^{\dag }a_{2}+ V_{12}a_{1}^{\dag }a_{2}+ V_{21}a_{2}^{\dag }a_{1}$
where  $V_{nm} = \sum_{ \alpha } \gamma^{~}_{  \alpha  n} \gamma^{*}_{ \alpha 
m}/{(\epsilon-\epsilon_  \alpha  )}$.
 
Taking matrix elements of Eq. \ref{eq:Tseries} between the graphene 2$p_z$
orbitals of the carbon atom(s) to which the adsorbed atom or molecule binds and
summing the resulting series yields
\begin{equation}
 \tilde{T} = \left( 1-\tilde{\mathcal{V}}\tilde{G}^0 
 \right)^{-1}\tilde{\mathcal{V}} 
 \label{eq:Tmatrix}
\end{equation}
where for the O atom adsorbate $\tilde{T}$, $\tilde{\mathcal{V}}$, $\tilde{G}^0$
and 1 are the 2$\times$2 matrices $\langle m|T|n \rangle$, $\langle
m|\mathcal{V}|n \rangle$, $\langle m|G^0|n \rangle$, and $\delta_{mn}$ with $|m
\rangle$ and $|n \rangle$ being the 2$p_z$ orbitals of the carbon atoms $m$ and
$n$ to which the O atom bonds. Here $m=1,2$ and $n=1,2$. For H, F and OH that
bond to one C atom (labelled 1) $\tilde{T}$, $\tilde{\mathcal{V}}$,
$\tilde{G}^0$ and 1 are the scalars $\langle 1|T|1 \rangle$, $\langle
1|\mathcal{V}|1 \rangle$, $\langle 1|G^0|1 \rangle$, and 1 respectively.

\section{Analytic Expressions for the Matrix Elements of $G^0$}
\label{analyticG}
The unperturbed Green's function matrix elements $\langle m|G^0|n \rangle =
G^0_{mn}$ between the relevant 2$p_z$ carbon orbitals of 2D graphene that enter
Eq. (\ref{eq:Tmatrix}) are given by
\begin{equation}
 G^0_{mn}(\epsilon) = \sum_{k,p} \frac{\langle m|\Phi_{kp}\rangle 
 \langle \Phi_{kp}|n\rangle}{\epsilon+i\eta- \langle 
 \Phi_{kp}|H_0|\Phi_{kp}\rangle}
 \label{eq:G0matrix}
\end{equation}
where $|\Phi_{kp}\rangle$ are the eigenstates of the unperturbed 
Hamiltonian $H_0$ of 2D graphene given by Eq.(\ref{pristine}), with wave 
vector $k$ and band index $p=\pm 1$.

We evaluate the diagonal matrix element  $\langle 1|G^0|1 \rangle =
G^0_{11}(\epsilon)$ as follows: For $m=n=1$ and small $\epsilon$ the summand in
Eq. \ref {eq:G0matrix} is strongly peaked in $k$-space around the Dirac points.
We therefore approximate the sum over the Brillouin zone in Eq. \ref
{eq:G0matrix} by the sum of integrals over two circles in $k$-space centered on
the two Dirac points $K$ and $K'$, choosing the area of each circle to be equal
to half of that of the Brillouin zone.  We then evaluate the integrals  by
linearizing $ \langle \Phi_{kp}|H_0|\Phi_{kp}\rangle$ in $k$ about each Dirac
point in the standard way which yields $ \langle \Phi_{kp}|H_0|\Phi_{kp}\rangle
\approx \pm 3 t \tau |k|/2$ where $\tau$ is the nearest neighbor spacing between
graphene carbon atoms.   This yields\cite{altderiv}
\begin{equation}\label{G11analytic } 
 G^0_{11}(\epsilon) \approx \frac{\epsilon}{\sqrt3 \pi t^2}
 ln\left(\frac{\epsilon^2}{\sqrt3\pi t^2 - \epsilon^2}\right) 
 - i\frac{|\epsilon|}{\sqrt3t^2},
\end{equation}
For comparison we carried out an exact numerical evaluation of Eq.
\ref{eq:G0matrix} {\em without} linearizing $ \langle
\Phi_{kp}|H_0|\Phi_{kp}\rangle$ or approximating the unperturbed model
Hamiltonian $H_0 = - \sum_{\left\langle n,m\right\rangle }t \left( a_{n}^{\dag
}a_{m}+h.c. \right)$ of 2D graphene or its eigenvalues or eigenfunctions in any
other way. We found the analytic approximation (Eq. \ref{G11analytic }) to agree
with our exact numerical results to within a few percent for small values of
$\epsilon$. We also found that the accuracy of Eq. (\ref{G11analytic }) can be
improved and extended to larger $|\epsilon|$ with the help of empirical
correction factors $\alpha (\epsilon) = 1.07(1+0.66 \epsilon^2/t^2)$ and $\beta
(\epsilon) = 1+0.31 \epsilon^2/t^2 + 0.33 \epsilon^4/t^4$. The resulting
expression
\begin{equation}\label{G11corrected } 
 G^0_{11}(\epsilon) = \frac{\epsilon \alpha (\epsilon)}{\sqrt3 \pi t^2}
 ln\left(\frac{\epsilon^2}{\sqrt3\pi t^2 - \epsilon^2}\right) - 
 i\frac{|\epsilon|\beta (\epsilon)}{\sqrt3t^2}
\end{equation}
is accurate in the range  $|\epsilon|/t \le 0.8$. We note that several less accurate 
analytic approximations for $G^0_{11}(\epsilon)$ have also been
proposed in the 
literature.\cite{Skrypnyk06, Skrypnyk07, Basko08, Pershoguba09,
Skrypnyk10} The first these\cite{Skrypnyk06} underestimated 
$G^0_{11}(\epsilon)$ by more than a factor of 2, due in part to the use of
an inadequate model of the graphene electronic structure with 
only 
a single Dirac cone instead of the two such cones that are centered at the 
Dirac 
points $K$ and $K'$ of graphene in reality. The most accurate of 
them\cite{Skrypnyk06, Skrypnyk07, Basko08, Pershoguba09,
Skrypnyk10}
is that given by Eq. 37a in Ref. \onlinecite{Basko08} with $\epsilon$
replaced by $|\epsilon|$ in the imaginary term. $G^0_{mn}(\epsilon)$ can also
be expressed exactly in terms of elliptic integrals that, however, must be 
evaluated numerically.\cite{Horiguchi}  

The analytic approximation scheme that leads to Eq. \ref{G11analytic } is
inappropriate for evaluating $G^0_{21}(\epsilon)$ directly because in that case
the summand in Eq. (\ref{eq:G0matrix}) is not maximal at the Dirac points $K$
and $K'$ where the linear approximation to $ \langle
\Phi_{kp}|H_0|\Phi_{kp}\rangle$ is accurate. However an analytic expression for 
$G^0_{21}(\epsilon)$ that is accurate in the same energy range as  Eq.
(\ref{G11corrected }) can still be obtained as follows.

We start from the identity 
\begin{equation}\label{GFandinverse} 
  \langle 1|(\epsilon+i\eta -H_0)G_0(\epsilon)| 1 \rangle = 1
\end{equation}
 from which it follows that
\begin{equation}\label{GFsum} 
 (\epsilon+i\eta)\langle 1|G_0(\epsilon)| 1 \rangle +\sum_{n = 2}^4 
 \langle 1| ( -H_0)| n \rangle\langle n|G_0(\epsilon)| 1 \rangle = 1
\end{equation}
In Eq. \ref  {GFandinverse} and \ref {GFsum},  $| 1 \rangle$ is the 2$p_z$
orbital of a carbon atom and $| n \rangle$ for $n=2,3,4$ are the 2$p_z$ orbitals
of its three nearest neighbors. Analysis of Eq. \ref {eq:G0matrix} shows that
because of the three-fold rotational symmetry of infinite 2D graphene $\langle
2|G_0(\epsilon)| 1 \rangle =\langle 3|G_0(\epsilon)| 1 \rangle = \langle
4|G_0(\epsilon)| 1 \rangle$. Then, since $\langle 1| ( -H_0)| n \rangle=t$ and
taking the limit $\eta \rightarrow 0$ we obtain from Eq. \ref {GFsum} the exact
result that
\begin{equation}\label{GFidentity} 
G^0_{21}(\epsilon) \equiv \langle 2|G_0(\epsilon)| 1 \rangle = \frac{1}{3t} 
- \frac{\epsilon}{3t} G^0_{11}(\epsilon)
\end{equation}
It is then straight forward to show also that $G^0_{21}(\epsilon) 
=G^0_{12}(\epsilon)$. Finally, inserting Eq. \ref {G11corrected }  
into Eq. \ref{GFidentity} we obtain
\begin{equation}\label{G21} 
G^0_{21}(\epsilon) =  \frac{1}{3t}-\frac{\epsilon^2 \alpha (\epsilon)}
{3\sqrt3 \pi t^3}ln\left(\frac{\epsilon^2}{\sqrt3\pi t^2 - \epsilon^2}\right) 
+ i\frac{\epsilon|\epsilon|\beta (\epsilon)}{3\sqrt3t^3}
\end{equation}
We have compared the analytic expression (\ref {G21}) with the results of our
exact numerical evaluation of $G^0_{21}(\epsilon)$ and found Eq. (\ref {G21}) to
be accurate under the same conditions as Eq. (\ref{G11corrected }), as expected.
Previous theoretical work\cite{Wehling07,Basko08} has only yielded analytic
expressions for $G^0_{21}(\epsilon)$ for the special case $\epsilon=0$. For that
case Eq. (\ref {G21}) reduces to $G^0_{21}(0) =  \frac{1}{3t}$ which agrees with
the result stated in Ref. \onlinecite{Basko08}  but differs in sign from that 
stated in Ref. \onlinecite{Wehling07}. It should be noted that the sign of 
$G^0_{21}$ is not arbitrary; the calculated Dirac point resonance for O 
adsorbed on graphene is modified significantly if this sign is reversed.

\section{Dirac Point Resonances of H, F, OH and O adsorbates on graphene}
\label{Dirac}

\begin{figure}[h]
\includegraphics[scale=0.4]{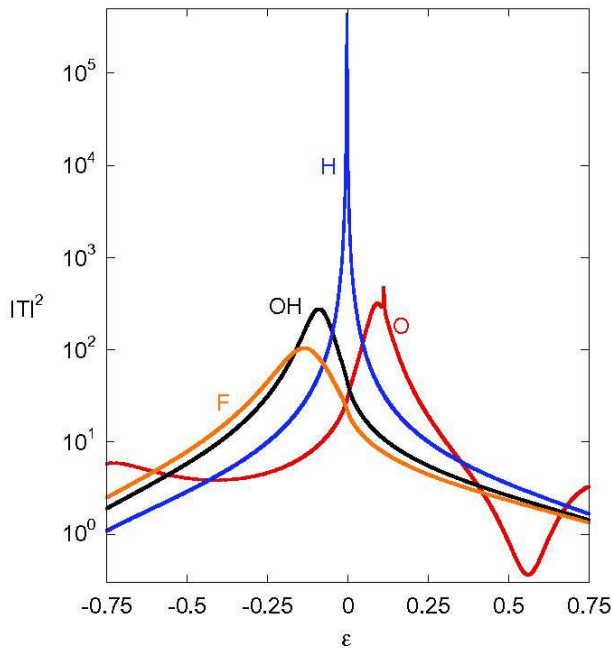}
\caption{(color online) Calculated square modulus of the $T$-matrix vs. electron
energy  $\epsilon$ for an H, F or O atom or OH group adsorbed on graphene in the
geometries shown in Fig. \ref{geom}. $T$ and $\epsilon$ are in units of
$t=2.7$eV. The Dirac point of graphene is at $\epsilon=0$. For  H, F and OH $T =
\langle 1|T|1 \rangle$. For O the square of the Frobenius norm of the matrix
$\langle m|T|n \rangle$ is plotted. The EMOs included in this calculation are
linear combinations of the atomic valence orbitals of the adsorbed species and
the 2$s$, 2$p_x$ and 2$p_y$ valence orbitals of each of the carbon atoms shown
in Fig. \ref{geom} for the respective adsorbed species. Thus the local
rehybridization of the graphene from the $sp^2$ to $sp^3$ bonding is included in
the model.The overlaps $ \sigma_{\alpha j}$ between the EMOs and the $2p_z$
orbitals of the carbon atoms  to which the adsorbed moieties bond are included
in the calculations.}
\label{resonanceplots}
\end{figure}
\begin{figure}[b]
\includegraphics[scale=0.4]{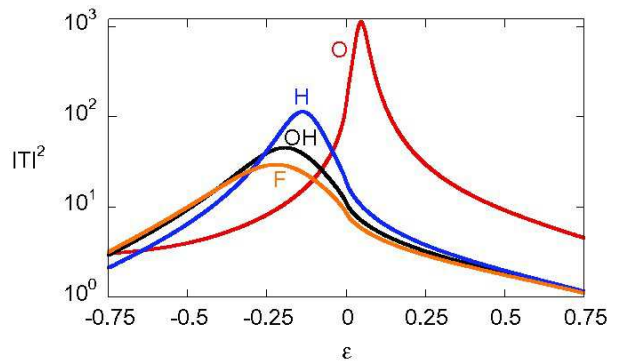}
\caption{(color online) Calculated square modulus of the $T$-matrix vs. electron
energy  $\epsilon$ for a simpler model of H, F or O atom or OH group adsorbed on
graphene than that used in the calculations presented in Fig.
\ref{resonanceplots}:  Here only the valence orbitals of the adsorbed species
themselves (no carbon orbitals) are included in the EMO's. The parameters
$\epsilon_\alpha$ and $\gamma_{\alpha j}$ used are from Table \ref{tab:2}; the
molecular orbitals $\psi_\alpha$ are used in the case of OH. The overlaps $
\sigma_{\alpha j}$ are also included in the calculation. Notation as in Fig.
\ref{resonanceplots}. }
\label{noCsigma}
\end{figure}

The strength of scattering associated with a defect is in general proportional
to the square modulus of appropriate matrix elements of the $T$-matrix. Thus the
energies $\epsilon$ at which resonant scattering by H, F, OH and O adsorbates
should occur are those at which  $|\langle m|T|n \rangle|^2$ have maxima. We
find these energies to be close to those at which
$|1-\tilde{\mathcal{V}}\tilde{G}^0|^{-2}$ for H, F and OH  or 
$|det(1-\tilde{\mathcal{V}}\tilde{G}^0)|^{-2}$ for O have maxima.

The square moduli of the matrix elements of the $T$-matrix defined in Section
\ref{Tmatrix} and calculated using the tight binding parameters $ \gamma_{\alpha
j}$ and $\epsilon_\alpha$ obtained from extended H\"{u}ckel theory as is
described in Section \ref{tbH} are shown vs. the electron energy $\epsilon$ in
Fig. \ref{resonanceplots}. The molecular orbitals $\psi_\alpha$ 
included in these calculations are linear combinations of the atomic valence
orbitals of the adsorbed species and the 2$s$, 2$p_x$ and 2$p_y$ valence
orbitals of all of the carbon atoms shown in Fig. \ref{geom} for the respective
adsorbed species. 13, 16, 17 and 22 EMOs are included in the calculations for 
H, F, OH and O respectively.  The overlaps $ \sigma_{\alpha j}$ between the EMOs
$\psi_\alpha$ and the $2p_z$ orbitals $\phi_j$ of carbon atoms $j$ to which the
adsorbed moieties bond are included in the calculations as is discussed at the
end of Section \ref{tbH}.

For each adsorbed species the $T$-matrix displays a prominent resonant peak (a
double peak for O in Fig. \ref{resonanceplots}) in the vicinity of $\epsilon=0$,
the Dirac point of graphene. The electron energy ${\epsilon_{\mathrm{DR}}}$ at
which the resonance is centered depends on the adsorbed species.
${\epsilon_{\mathrm{DR}}}=  -0.136t,  -0.089t, -0.0026t$ for F, OH and H,
respectively. For O there is a narrow peak near  $0.112t$ that overlaps a
broader peak centered near $0.090t$. The strengths of the F, OH and broader O
resonances (as measured by the area under the resonance curve when plotted on a
linear scale) are all comparable. The narrow O resonance is an order of
magnitude weaker than these while the H resonance is two orders of magnitude
stronger.

For comparison the results of a similar calculation but for a simpler model in
which the carbon atom 2$s$, 2$p_x$ and 2$p_y$ valence orbitals are omitted from
the EMOs are shown in Fig. \ref{noCsigma}. In this case
${\epsilon_{\mathrm{DR}}}=  -0.222t,  -0.194t,  -0.138t$ and $   0.046t$ for F,
OH, H and O, respectively.\cite{no_overlap} Comparing  Fig. \ref{resonanceplots}
with Fig. \ref{noCsigma} it is evident that the coupling of the adsorbate to the
graphene  carbon 2$s$, 2$p_x$ and 2$p_y$ valence orbitals (that are involved in
the partial rehybridization of the carbon atom to which the adsorbate bonds to
the $sp^3$ electronic structure) can affect adsorbate-induced Dirac point
resonances very strongly: It  is directly responsible for the H resonance in
Fig. \ref{resonanceplots} being two orders of magnitude stronger than the
resonances for F, O and OH. It is also responsible for the {\em double} peak
structure of the O resonance in Fig. \ref{resonanceplots} that is absent in 
Fig. \ref{noCsigma}. Notice also the {\em antiresonance} in the O $T$-matrix
near  $\epsilon = 0.55t$ in Fig. \ref{resonanceplots} that is absent in Fig.
\ref{noCsigma}.

The Dirac point resonance energy increases from F to OH to H to O in both models
and the sign of the resonance energy for each species (negative for  F, OH and H
and positive for O)  is the same in both models. However, the coupling of the H
adsorbate to the graphene  carbon 2$s$, 2$p_x$ and 2$p_y$ valence orbitals
results in the H Dirac point resonance being extremely close to the Dirac point
of graphene  (${\epsilon_{\mathrm{DR}}}= -0.0026t$) in Fig.
\ref{resonanceplots}. As will be seen in Section \ref{results} this results in
the conductances of graphene nanoribbons with hydrogen adsorbates being almost
symmetric about the graphene Dirac point (as they are for ribbons with carbon
atom vacancies\cite{disorder09}), in marked contrast to the asymmetric
conductances for the other adsorbed species.

\begin{table}[t]
\caption{Minimal set of effective tight-binding parameters $\epsilon_\alpha$
and$\gamma_{\alpha j}$ in units of $t=2.7$ eV for adsorbed H, F and O and OH on
graphene. The EMO energies $\epsilon_\alpha$ are measured from the Dirac point
energy of graphene. $\pm$ means that $\gamma_{\alpha j}$ has opposite signs for
the two carbon atoms to which the O atom bonds.
}
\begin{center}
\begin{tabular}{lc|c|c|c}
 \multicolumn{2}{l|}{adsorbate} & $\epsilon_\alpha$ & $\gamma_{\alpha j}$ & \\ 
 \hline
 H & $ $ & -0.0383 & 2.219 &  \\
 \hline
 F & $ $ & -10.862  &  4.363  &   \\
    & $ $ & -2.460  &  1.645  &    \\
    & $ $ & -0.914  &  1.180  &    \\
 \hline
 OH & $ $ & -8.536 &   3.203  &    \\
    & $ $ & -1.820  &  1.779   &   \\
    & $ $ & -0.709 &  1.540   &   \\
 \hline
 O & $ $ & -5.356 &     3.240 & \\
   & $ $ & -1.448 &   $\pm 1.000$ & \\
   & $ $ & -0.373 &    1.095 & \\
   & $ $ & 0.130 & $\pm  0.176$ & \\
   & $ $ & 1.463 &  1.650 & \\
\end{tabular}
\end{center}
\label{tab:1}
\end{table}

Finally we find that an accurate description of the Dirac resonance profiles in
Fig. \ref{resonanceplots} (including their energies, widths and heights) can be
obtained by including in the tight binding Hamiltonian Eq.
(\ref{eq:hamiltonian}) relatively small sets of EMOs since some of the EMOs
couple only weakly to the graphene $\pi$ system. The EMO energies
$\epsilon_\alpha$ and coupling parameters  $\gamma_{\alpha j}$ for a minimal
tight binding model Hamiltonian that describes the Dirac point resonance
profiles shown in Fig. \ref{resonanceplots} are presented in Table \ref{tab:1}.
The EMO energy and coupling parameter sets presented in Table \ref{tab:1} 
will be used in the more sophisticated 
version of our transport
calculations on graphene nanoribbons that we report in Section \ref{results}.
The values of $\epsilon_\alpha$ given in Table \ref{tab:1} are the energies of
EMOs calculated from extended H\"{u}ckel theory as discussed in Section
\ref{tbH}. However the values of some of the $\gamma_{\alpha j}$ that are given
have been adjusted so that the small set of parameters $\epsilon_\alpha$ and
$\gamma_{\alpha j}$ given in Table \ref{tab:1} yields a good fit to the
resonance profiles in the energy range $-0.75t < \epsilon < 0.75t$ shown in 
Fig. \ref{resonanceplots} without the need to include the overlaps $
\sigma_{\alpha j}$ in the calculation.

\section{Comparison with Density Functional Theory-Based Models of Dirac 
Point Resonances}
\label{DFTvsEH}

In the previous attempts to construct tight binding models of Dirac point
resonances for use in transport calculations in graphene, the tight-binding
parameters were obtained by fitting the results of {\em ab initio} density
functional theory-based electronic structure calculations to very simple tight
binding models. In those models the adsorbed atom or molecule M was described by
just a {\em single} effective orbital energy parameter  $\epsilon_{\mathrm{M}}$
and a {\em single} coupling parameter $\gamma_{\mathrm{M}}$. For example,
$\epsilon_{\mathrm{H}} = 0.66 t$ and $\gamma_{\mathrm{H}} = 0.22 t$ were
obtained for adsorbed hydrogen and $\epsilon_{\mathrm{OH}} = -2.9 t$ and
$\gamma_{\mathrm{OH}} = 2.3 t$ were obtained for the adsorbed hydroxyl group in
Ref. \onlinecite{Robinson08}. On the other hand, in Ref. \onlinecite{Wehling10}
{\em much smaller} values of the effective orbital energy parameter
$\epsilon_{\mathrm{M}} \le 0.1 t$  were found for a number of covalently bonded
adsorbed species, including hydrogen, and $\gamma_{\mathrm{M}} \ge 2 t$ was
found for the same species.

That such very different results have been obtained from density functional
theory-based calculations even for hydrogen, the simplest of all adsorbed
species, raises the question whether density functional calculations, although
putatively a ``first principles" method, are a sound basis for theoretical
studies of the Dirac point resonances of graphene with adsorbates. This
demonstrated lack of consistency may be related to the fact that density
functional theory, although well suited to calculations of the total ground
state energies of many condensed matter systems, is known to have important {\em
fundamental} deficiencies as a methodology for electronic structure
calculations; for a recent discussion of the relevant physics and a review of
the literature the reader is referred to Ref. \onlinecite{Kirczenow}.

As a consequence of these deficiences, for example, density functional
calculations underestimate the band gap of silicon and other semiconductors by
as much as a factor of two. They also yield offsets between the energy levels of
molecules adsorbed on silicon and the silicon valence band edge that more
sophisticated theories indicate to be in error by as much as
1.4eV.\cite{Quek2007} The latter error is similar in size to the above-mentioned
discrepancy between the values of $\epsilon_{\mathrm{H}}$ predicted by the
density functional theory calculations reported in Refs. \onlinecite{Robinson08}
and \onlinecite{Wehling10} for H adsorbed on graphene.

Because different density functional theory-based electronic structure
calculations have yielded such different results even for the Dirac point
resonance due to hydrogen on graphene, we chose instead to base our electronic
structure calculations on the well known {\em semi-empirical} extended
H\"{u}ckel model of quantum chemistry.  The extended H\"{u}ckel model has been
parameterized\cite{yaehmop} based on a large body of {\em experimental}
electronic structure data for atoms and molecules and has been used successfully
in electronic transport calculations for a variety of molecular systems, as has
been outlined in Section \ref{Introduction}. The relative merits of electronic
structure calculations based on density functional theory and those based on
extended H\"{u}ckel theory have been discussed in detail in Section 4.7 of Ref.
\onlinecite{Kirczenow}.

The minimal tight-binding model parameters for the Dirac point resonance due to
hydrogen adsorbed on graphene that we derived from extended H\"{u}ckel theory
using the methodology described in Sections \ref{tbH}-\ref{Dirac} are given in
Table \ref{tab:1}. They are in excellent agreement with the corresponding
density functional theory-based results $\epsilon_{\mathrm{M}} \le 0.1 t$ and
$\gamma_{\mathrm{M}} \ge 2 t$ stated in Ref. \onlinecite{Wehling10} but are not
consistent with the density functional theory-based results
$\epsilon_{\mathrm{H}} = 0.66 t$ and $\gamma_{\mathrm{H}} = 0.22 t$ reported in
Ref. \onlinecite{Robinson08}.

We find {\em one}-orbital tight binding models such as those used in Refs.
\onlinecite{Robinson08} and \onlinecite{Wehling10} not to yield a satisfactory
description of Dirac point resonances for the other adsorbed species (F, OH and
O) that we have considered within extended H\"{u}ckel theory. Therefore
comparing our tight-binding models for the Dirac point resonances of those
species directly with the corresponding tight binding parameters that have been
derived from density functional theory (as we have done above for the case of
adsorbed hydrogen) is not possible. However, we have calculated the local
densities of states (LDOS) associated with the Dirac point resonances within our
extended H\"{u}ckel based model for these adsorbates and compare them below with
the corresponding LDOS features calculated using density functional theory in
Ref. \onlinecite{Wehling09}.

\begin{figure*}[t]
\includegraphics[keepaspectratio,width=\textwidth]{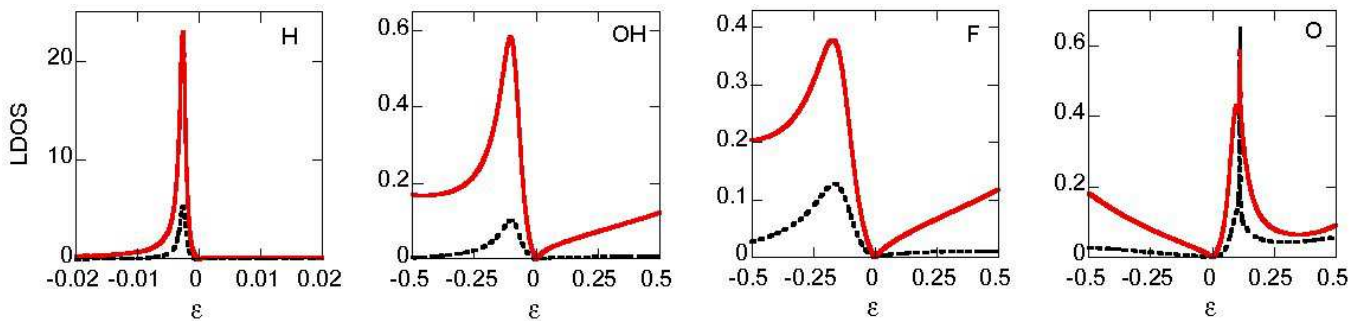}
\caption{(Color online) Calculated local densities of states (LDOS) vs. electron
energy  $\epsilon$ for an H, F or O atom or OH group adsorbed on graphene in the
geometries shown in Fig. \ref{geom}. The LDOS is in units of $1/t$ and
$\epsilon$ is in units of $t=2.7$eV. For F, O atom OH the black dashed curves
show the LDOS associated with the 2$p_z$ orbital of a carbon atom to which the
adsorbed moiety bonds. For H the black dashed curve shows the LDOS associated
with the 2$p_z$ orbital of a carbon atom to which the adsorbed atom bonds
multiplied by 500. The red solid curves show the LDOS associated with the 2$p_z$
orbital of a carbon atom that is a nearest neighbor of a carbon atom to which
the adsorbed moiety bonds. The Dirac point of graphene is at $\epsilon=0$. The
EMOs included in this calculation are linear combinations of the atomic valence
orbitals of the adsorbed species and the 2$s$, 2$p_x$ and 2$p_y$ valence
orbitals of each of the carbon atoms shown in Fig. \ref{geom} for the respective
adsorbed species. Thus the local rehybridization of the graphene from the $sp^2$
to $sp^3$ bonding is included in the model.}
\label{fig:4}
\end{figure*}

Our calculated local densities of states for graphene with an adsorbed H, F or O
atom or OH group are shown in Fig. \ref{fig:4}. The extended H\"{u}ckel
theory-based model used in these calculations is the same as that used to
calculate the square moduli of the $T$-matrices that are shown in Fig.
\ref{resonanceplots}. That is, the EMOs included the calculations are linear
combinations of the atomic valence orbitals of the adsorbed species and the
2$s$, 2$p_x$ and 2$p_y$ valence orbitals of each of the carbon atoms shown in
Fig. \ref{geom} for the respective adsorbed species. Thus the model includes the
effects of the local rehybridization of the graphene due to the presence of the
adsorbate. The quantities plotted in Fig. \ref{fig:4} are the partial LDOS
defined by $D_n(\epsilon)=-\mathrm{Im}(\langle n|G(\epsilon)|n\rangle)/\pi$
where  $G$ is defined by Eq. \ref{eq:G} and $|n\rangle$ represents a 2$p_z$
orbital of a carbon atom. The dashed black curves are for graphene carbon atoms
to which the adsorbed moieties bond and the solid red curves are for nearest
(carbon) neighbors of those carbon atoms. Note the different energy scale used
for the case of H and also that the dashed black curve for H has been scaled up
for clarity by a factor of 500. In each plot the Dirac point is at $\epsilon=0$.
The energies at which the peaks of the densities of states occur in Fig. 
\ref{fig:4} are reasonably close to the energies of the Dirac point resonance
peaks for the respective adsorbates in Fig. \ref{resonanceplots}, the largest
discrepancy being $\sim 20$\% for fluorine.

The partial LDOS in Fig. \ref{fig:4} for H, F and O can be compared with the
corresponding density functional theory-based results in Figs. 1(c), 1(d) and 
5(a) of Ref. \onlinecite{Wehling09}, respectively. However, some differences
between the systems considered should be noted: The results in Fig. \ref{fig:4}
are for a single adsorbed atom or molecule on an infinite graphene sheet 
whereas those in Ref. \onlinecite{Wehling09} are for periodic structures with $4
\times 4$ graphene supercells each containing an adsorbed moiety.  Thus in Ref.
\onlinecite{Wehling09}, in contrast to the present work, the Dirac point
resonance is expected to be broadened due to the presence of multiple adsorbed
atoms on the graphene. Also, the DOS features for  H, F and O in Ref.
\onlinecite{Wehling09} are located relative to the Fermi energy. The latter may
be close to the Dirac point energy but its location relative to the Dirac point
energy is not determined {\em precisely} since in the model systems studied in
Ref. \onlinecite{Wehling09} there are no very large regions of pristine graphene
with no adsorbate where the Dirac point is well defined.

With these caveats, the LDOS for the Dirac point resonance for H in Fig.
\ref{fig:4} is consistent with that in Ref. \onlinecite{Wehling09}, although for
the latter the resonance is broader and the partial LDOS on the carbon atom to
which the H bonds is not as weak relative to the partial LDOS on its nearest
carbon atom neighbors. The LDOS for the fluorine Dirac point resonance in Ref.
\onlinecite{Wehling09} is also similar to that in Fig.4 although the LDOS peaks
in the latter are somewhat lower in energy relative to the Dirac point than
those in the former are relative to the Fermi energy. However, as we have
already noted, the precise location of the Fermi level relative to the Dirac
point is uncertain in the density functional theory-based calculations.

The LDOS for O in Fig. \ref{fig:4} differs markedly from that in Ref.
\onlinecite{Wehling09}: There is no obvious peak in the LDOS in the immediate
vicinity of the Fermi energy in the latter case in contrast to the peaks
associated with the Dirac point resonance $\sim 0.1t$ above the Dirac point in
the former. The reasons for this difference are not clear at present. We note,
however,  that, as has been discussed above, different density functional theory
based calculations\cite{Robinson08, Wehling10} have yielded very different
results even for the Dirac point resonance for adsorbed hydrogen. Furthermore,
recent theoretical work\cite{Ramasubramaniam2010} has shown even {\em gross}
features of the electronic structures of narrow graphene nanoribbons with high
concentrations of adsorbed O at the ribbon edges (including the presence or
absence of a large band gap at the Fermi level) calculated using density
functional theory to be sensitive to the precise choice of the
exchange-correlation energy functional used in the calculations. While the
calculated LDOS for O in Ref. \onlinecite{Wehling09} resembles qualitatively
that found for a ``double impurity" in a simple tight binding
model\cite{Wehling09} it should be noted that in that model the ``double
impurity" represents {\em two} atoms that are adsorbed on adjacent carbon atoms
and do not interact with each other directly, a situation that is very different
than a {\em single} oxygen atom that bonds to {\em two} adjacent carbon atoms in
its lowest energy configuration in reality and in our extended H\"{u}ckel
theory-based model.

Given the differing theoretical predictions that we have discussed above, it is
evident that experiments probing adsorbate-induced Dirac point resonances in
graphene would be of considerable interest.

\section{Model Hamiltonian for Transport Calculations in Graphene Nanoribbons}
\label{model}

In this Section we describe how the tight-binding Hamiltonians developed in the
preceding Sections for adsorbates on infinite 2D graphene are adapted for
calculations of quantum transport in graphene nanoribbons with adsorbates that
we consider in the remainder of this paper.

We describe the graphene ribbons by the tight-binding Hamiltonian
\begin{equation}
 H= H_{\pi} + \sum_{\alpha}\epsilon_{\alpha}d_{\alpha}^{\dag }d_{\alpha} + 
 \sum_{\alpha, j} \gamma_{\alpha j} \left( d_{\alpha}^{\dag }a_{j}+h.c. 
 \right)
 \label{eq:tbhamiltonian}
\end{equation}%
where 
\begin{equation}
  H_{\pi}= - \sum_{\left\langle i,j\right\rangle }t_{ij}
\left( a_{i}^{\dag }a_{j}+h.c. \right).
 \label{eq:pihamiltonian}
\end{equation}%
$t_{ij}$ is the Hamiltonian matrix element between nearest-neighbor $2p_z$
carbon orbitals of the $\pi$ band of the graphene nanoribbon.  $a_{i}^{\dag }$
is the creation operator for an electron in $2p_z$ carbon orbital $i$.
$d_{\alpha}^{\dag }$ creates an electron in an extended molecular orbital (EMO)
$\psi_\alpha$ that is associated with an adsorbed atom or molecule and has
energy $\epsilon_{\alpha}$. As defined in Section \ref{tbH}, an EMO is a linear
combination of the valence orbitals of the adsorbed atom or molecule and (in the
more sophisticated versions of the model) the $2s$, $2p_x$ and $2p_y$ valence
orbitals of the graphene carbon atom(s) to which the adsorbate bonds and
neighboring graphene carbon atoms. The graphene $2p_z$ orbitals are not included
in the EMOs since they are included  in the tight binding Hamiltonian $H$
through $H_{\pi}$. In the present work we will include for simplicity only
coupling matrix elements  $ \gamma_{\alpha j}$ between the EMOs associated with
the adorbates and the $2p_z$ valence oribitals of the graphene carbon atoms to
which that adsorbed atoms or molecules bond.

In the graphene ribbon $\pi$-band Hamiltonian $H_{\pi}$ we include nearest
neighbor Hamiltonian matrix elements $t_{ij}$. For most of these we set
$t_{ij}=t=2.7$eV, the usual value for tight binding theories of pristine
graphene.\cite{Reich02, review} However, interaction with the H, F, OH and O
adsorbates shifts the carbon atoms to which these moieties bond out of the
graphene plane by fractions of an Angstrom, perturbing the values of $t_{ij}$
between the carbon atoms to which the adsorbates bond and their neighbors.
Although we find the effect of this change in $t_{ij}$ on electron transport in
ribbons to be modest (typically less that a 15\% difference in the conductance) 
we include it in the calculations presented in this article, estimating the
modified values of $t_{ij}$ by applying extended H\"{u}ckel theory to the
relaxed geometries of the graphene in the presence of the
adsorbates.\cite{scaling} We do not, however, consider edge reconstruction
effects and spin and electron interaction phenomena;  these are outside of the
scope of the present study.

An important effect associated with the shifting of carbon atoms out of the
graphene plane that occurs upon adsorption of H, F, OH or O is the local partial
rehybridization of the graphene from $sp^2$ to $sp^3$ bonding. We shall
elucidate the role that the rehybridization plays in transport in graphene
ribbons by comparing the results of transport calculations for two models:

(i) A simpler model in which the EMOs associated with adsorbed atoms or
molecules are approximated by linear combinations of only the valence orbitals
of the adsorbed species.

(ii) A model in which the EMOs are linear combinations of the the valence
orbitals of the adsorbed species {\em and} of the appropriate valence orbitals
($2s$, $2p_x$ and $2p_y$) of the graphene carbon atoms involved in the
rehybridization, i.e., the carbon atoms to which the adsorbate bonds and their
nearest neighbors.

For case (ii) the number of EMOs per adsorbed moiety is too large for exact
quantum transport calculations to be carried out with the computational
resources available to us for ribbons of experimentally relevant sizes and
adsorbate concentrations of interest. However it was found in Section
\ref{Dirac}  that reduced sets of between 1 and 5 EMOs per adsorbed moiety with
suitably adjusted values of $ \gamma_{\alpha j}$ are sufficient to provide an
accurate description of the Dirac point resonances in graphene induced by H, F,
OH and O adsorbates. These minimal sets (with parameters listed in Table
\ref{tab:1}) will be used in the transport calculations for case (ii) that we
present in this paper.

In our transport calculations the adsorbed atoms and molecules are introduced by
randomly placing them on the graphene surface. They are characterized by the
probability $p$ to find an adsorbed moiety per  carbon atom.  In order to
convert $p$ to the usual concentration one should scale it by the number of
carbon atoms divided by the sample size, $3.8\times 10^{19}$ m$^{-2}$.

 In the linear response regime the zero temperature conductance of the 
 graphene ribbon is given by the Landauer 
 formula\cite{Landauer70, Economou81, Fisher81,Buttiker,Kirczenow}
\begin{equation}
	G = \frac{2e^{2}}{h} \sum_{ij} T_{ji}.
	\label{eq:conductance}
\end{equation}
where $T_{ji}$ is the transmission coefficient from the subband $i$ in 
the left lead to the subband $j$ the right lead, at the Fermi energy. 
For non-zero temperatures  
\begin{equation}
	G = -\frac{2e^{2}}{h} \int_{-\infty}^{\infty }dE \;T(E) 
	\frac{\partial f(E)}{\partial E}
	\label{eq:ftconductance}
\end{equation}
where
$T(E) = \sum_{ij}T_{ji}(E)$ and $f(E)$ is the Fermi distribution 
function. 
$T_{ji}$ is calculated by the recursive Green's function method, 
see Ref. \onlinecite{Igor08} for details. 

\section{Results}
\label{results}

\begin{table}[t]
\caption{Tight-binding parameters for the adsorbed H, F, O and OH in units of
$t=2.7$ eV for the simplest model in which the EMOs do not include the carbon
$2s$, $2p_x$ and $2p_y$ atomic orbitals. The atomic and molecular orbital
energies $\epsilon_\alpha$ are measured from the Dirac point of graphene. 
$\gamma_{\alpha, C_{2p_z}} $ is the Hamiltonian matrix element between the
orbital $\psi_\alpha$ of the adsorbate and the $2p_z$ orbital of the nearest
graphene C atom. The $\pm$ means that the $ \gamma_{\alpha, C_{2p_z}} $ values
for the two C atoms to which the O atom bonds have opposite signs. For OH the
parameters are given for both the atomic O and H orbitals and the molecular
orbitals $\psi_\alpha$ of OH. $\gamma_{C_{2p_z},C_{2p_z}}$ is the Hamiltonian
matrix element between the $2p_z$ orbital of the graphene C atom to which the
adsorbed moiety bonds and the $2p_z$ orbital of a neighboring C
atom.\cite{scaling} For O the first $\gamma_{C_{2p_z},C_{2p_z}}$ is for the pair
of C atoms to which the O bonds while the second is for a C atom to which the O
bonds and another nearest C neighbor of that C atom.
}
\begin{center}
\begin{tabular}{lc|c|c|c}
 adsorbate & $\psi_\alpha$ & $\epsilon_\alpha$ & $ \gamma_{\alpha, 
 C_{2p_z}} $ & $\gamma_{C_{2p_z},C_{2p_z}}$\\ 
 \hline
 H & $1s$ & -0.81 & 1.89 & 0.79\\
 \hline
 F & \begin{tabular}{l}$2s$\\$2p_z$\end{tabular} & 
 \begin{tabular}{c}-10.59\\-2.48\end{tabular} & 
 \begin{tabular}{c}4.70\\1.45\end{tabular} & 0.79 \\
 \hline
 OH & \begin{tabular}{l}$2s^O$\\$2p_z^O$\\$1s^H$\end{tabular} & 
 \begin{tabular}{c}-7.74\\-1.26\\-0.81\end{tabular} & 
 \begin{tabular}{c}4.10\\1.24\\0.36\end{tabular} & 0.73 \\
  \hline
 OH & \begin{tabular}{l}$\psi_1$\\$\psi_2$\\$\psi_3$\end{tabular} & 
 \begin{tabular}{c}-8.17\\-1.64\\7.39\end{tabular} & 
 \begin{tabular}{c}3.75\\1.81\\1.69\end{tabular} & 0.73 \\
 \hline
 O & \begin{tabular}{l}$2s$\\$2p_z$\\$2p_x$\end{tabular} & 
 \begin{tabular}{c}-7.74\\-1.26\\-1.26\end{tabular} & 
 \begin{tabular}{c}3.47\\0.76\\$\pm 0.80$\end{tabular} & 
 \begin{tabular}{c}0.92 \\ 0.89\end{tabular}
\end{tabular}
\end{center}
\label{tab:2}
\end{table}

\begin{figure*}[t]
\includegraphics[keepaspectratio,width=\textwidth]{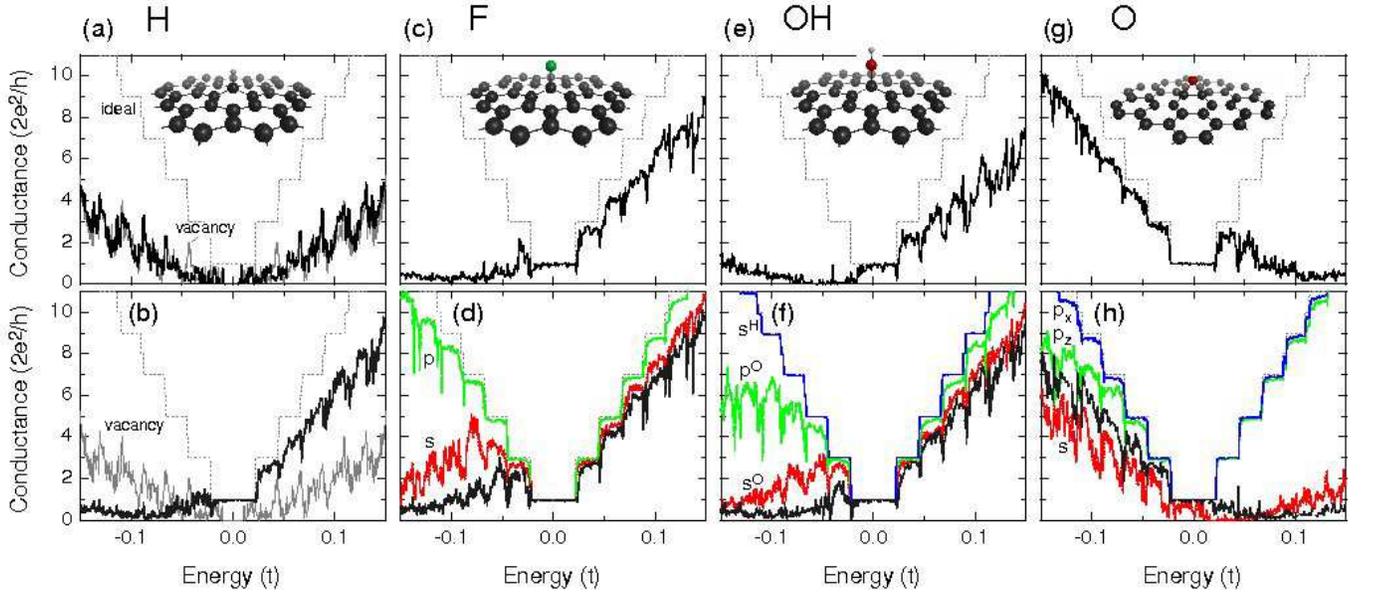}
\caption{(color online) Calculated conductances as a function of the Fermi
energy for graphene ribbons with different adsorbed species at a concentration
$p=10^{-4}$ .  The relaxed geometries of the adsorbed species and nearby
graphene atoms are shown in the insets. The dotted grey lines show the
conductance of the ideal ribbon without any defects. The solid black lines in
(a), (c), (e) and (g) show the calculated conductances of ribbons with H, F, OH
and O adsorbates respectively for the model that includes the effects of the
adsorbate induced rehybridization of the graphene; the parameterization used is
given in Table  \ref{tab:1}. The solid black lines in (b), (d), (f) and (h) show
the calculated conductances of ribbons with H, F, OH and O adsorbates
respectively for the simpler model that includes only the adsorbate valence
orbitals in the EMOs as parameterized in Table \ref{tab:2} and thus does not
include the effects of the rehybridization of the graphene. Red, green and blue
solid lines show the effect on the conductance of the individual orbitals of the
adsorbed species as indicated. The grey solid line in (a) and (b) shows the
conductance of a ribbon with interior carbon atom vacancies at
$p_{vac}=10^{-4}$.  Ribbon width $W=30$ nm; length $L=500$ nm. Temperature
$T=0$. $t=2.7$ eV.}
\label{fig:5}
\end{figure*}

\begin{figure}[t]
\includegraphics[keepaspectratio,width=\columnwidth]{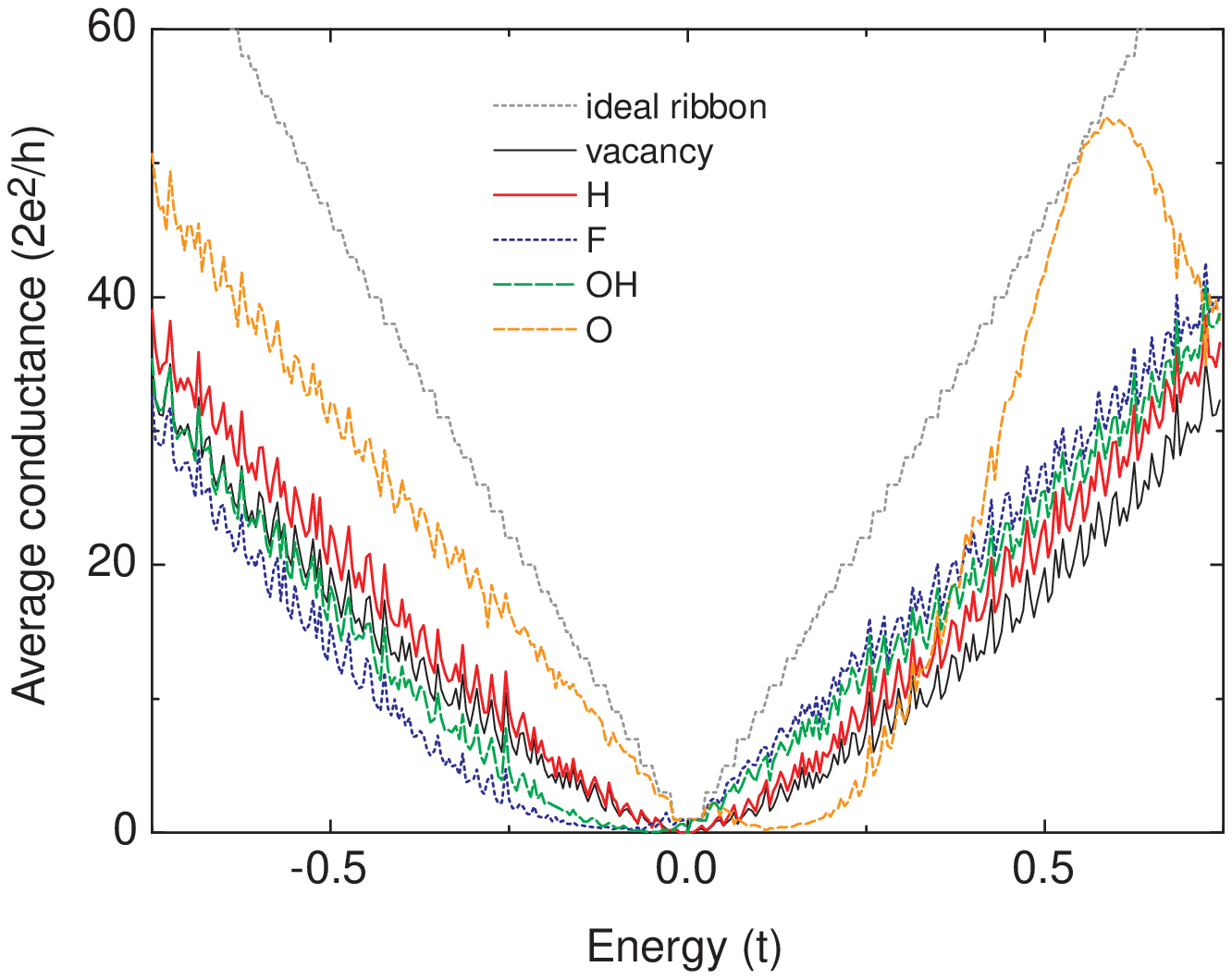}
\caption{(color online) Comparison of the calculated averaged conductances of
graphene ribbons with adsorbates (for the model that includes adsorbate-induced
rehybridization of the graphene) and interior carbon atom vacancies at a
concentration $p=10^{-4}$ in each case. The averaging in each plot is over 10
arrangements of the positions of the vacancies or adsorbed atoms or molecules.
The grey dotted line shows the conductance of the ideal ribbon without any
defects. }
\label{fig:6}
\end{figure}

In our transport calculations the ribbon has a width $W=30$nm as in the
experiments of Lin et al.\cite{Lin08} The adsorbed H, F, OH or O are assumed to
be present only in a finite region of  length $L=500$nm of the ribbon which is
attached at its two ends to semi-infinite leads represented by ideal ribbons of
the same width. The edge configuration is taken as armchair in the following.
For comparison we  also present some results for ribbons with interior carbon
atom vacancy defects.\cite{disorder09}

\subsection{Ribbons with H Adatoms}
\label{H_Adatoms}

\subsubsection{Atomic Geometry and Electronic Structure}

Hydrogen is the simplest adsorbate that bonds covalently to a carbon atom of the
graphene lattice. According to the extended H\"{u}ckel model, the H $1s$ orbital
energy locates not far from the Dirac point of graphene,
$\epsilon_{H_{1s}}=-0.81t$, and the Hamiltonian matrix element between the
hydrogen $1s$ orbital and the carbon $2p_z$ orbital of the closest graphene
carbon atom, $ \gamma_{H_{1s}, C_{2p_z}}= 1.89t$ (see Table \ref{tab:2}), is
nearly twice as large as the Hamiltonian matrix element $t$ between the $2p_z$
orbitals of adjacent carbon atoms. Adsorption of the H atom also results in a
modified graphene lattice geometry. In the relaxed geometry, we find significant
 lifting of the carbon atom directly bound to the H out of the graphene plane by
0.35\AA. This is accompanied by weakening of the $C_{2p_z} - C_{2p_z}$
Hamiltonian matrix element between the C atom to which the H atom binds and its
neighbor C atoms to $0.79t$. Since the graphene sheet is no longer planar,
partial rehybridization from $sp^2$ to $sp^3$ occurs near the adsorbed H atom:
The carbon atom to which the H atom binds can be regarded as forming $\sigma$
bonds with its carbon atom neighbors and with the H atom.  The rehybridization
of the graphene on adsorption of the H atom and the strong coupling between H
adatom and the graphene result in strong scattering of graphene $\pi$-band
electrons near the adsorbed H atom.

\subsubsection{Conduction in Ribbons with H Adatoms}
\label{H_conductance}

In Fig \ref{fig:5} (a) and (b) we show the calculated conductance of a graphene
nanoribbon with a concentration $p^H=10^{-4}$ of adsorbed H. In Fig \ref{fig:5}
(a) the results are shown for the model that includes in the adsorbate EMO the H
$1s$ orbital and the $2s$, $2p_x$ and $2p_y$ orbitals of the carbon atom to
which the H binds and its neighboring carbon atoms, the tight-binding parameters
used being those given in Table  \ref{tab:1}. This model includes the effect of
the local rehybridization of the graphene from $sp^2$ to $sp^3$ bonding. For
comparison the results for a model in which the adsorbate EMO includes only the
H $1s$ orbital (with the parameters given in Table \ref{tab:2} of the present
paper) are shown in Fig \ref{fig:5}(b). The conductances of the same ribbon with
no adsorbate but an equal concentration $p=10^{-4}$ of interior carbon atom
vacancies\cite{disorder09} and of an ideal ribbon with no adsorbate or vacancies
are also shown. In both models even for this low concentration of adsorbed H
atoms the conductance of the ribbon is strongly suppressed relative to that of
the ideal ribbon.

The calculated conductance with the rehybridization of the graphene taken into
account (the solid black curve in Fig \ref{fig:5} (a)) is strikingly similar
both qualitatively and quantitatively to that of the ribbon with the same
concentration of carbon atom vacancies (the solid grey curve), whereas the
conductance calculated without including rehybridization (the solid black curve
in Fig \ref{fig:5} (b)) is qualitatively different.

This is consistent with the idea\cite{Wehling09X} that the rehybridization
should effectively decouple the carbon atom to which the H atom bonds from the
graphene $\pi$ band, in which case scattering of graphene $\pi$ band electrons
by an adsorbed H atom would be expected to resemble electron scattering by a
carbon atom vacancy. However this simple picture does not account for the
differences between graphene with a H adsorbate and graphene with a F or OH
adsorbate: As will be seen below the  conductance characteristics of ribbons
with F and OH adsorbates with rehybridization included differ qualitatively from
those of ribbons with vacancies and change much less drastically than those for
H when rehybridization is included in the model, although the changes in the
graphene geometry due to the adsorption of F and OH are very similar to and even
slightly larger than for H adsorption.

If the energy scale is broadened (see Fig. \ref{fig:6}) an approximately linear
increase of the conductance with the absolute value of the energy is found with
nearly the same absolute slope at positive and negative energies for ribbons
with adsorbed H. Thus on the larger energy scale the conductance of the ribbon
is nearly symmetric in the Fermi energy in presence of adsorbed H as it is in
the presence of interior carbon atom vacancies. This behavior of the conductance
is consistent with experiment.\cite{Lin08}

\subsubsection{Role of the Dirac Point Resonance}

A clearer understanding of the role of the graphene rehybridization in electron
transport can be gained by comparing the the conductance plots in  Fig
\ref{fig:5} (a) (where the rehybridization is included in the model) and Fig
\ref{fig:5} (b) (where it is not) with the calculated properties of the Dirac
point scattering resonances associated with adsorbed H in the same two models as
were used for these transport calculations. The corresponding Dirac point
resonances are shown in Fig.  \ref{resonanceplots} and  \ref{noCsigma}
respectively. Ignoring the mesoscopic conductance fluctuations in Fig.
\ref{fig:5}, it is apparent that including the rebybridization in the transport
model results in a shift in the energy of the main conductance minimum from
$\sim -0.1t$ in Fig.\ref{fig:5}(b) to $\sim 0.0t$ (the Dirac point of graphene)
in Fig.\ref{fig:5}(a). This matches reasonably well the shift in the energy of
the hydrogen Dirac point  scattering resonance from $-0.138t$ to $-0.0026t$ that
was found upon inclusion of rehybridization in the theory of the H Dirac point 
resonance in Section \ref{Dirac}. Thus the locations in energy of the
conductance minima for the two models agree quite well with the energies at
which the Dirac point resonances of H occur in the models. This is reasonable
since strong resonant electron scattering near a particular energy is expected
to suppress electron transport near that energy. As will be seen below there is
similar agreement between the Dirac point resonance energies and the energies at
which the nanoribbon conductance minima occur for the other adsorbates that we
study in this article.

A notable feature of the conductance calculated for the nanoribbons with
adsorbed hydrogen in the model that does {\em not} include the graphene
rehybridization (Fig.\ref{fig:5}(b)) is that in this case the conductance is
affected only weakly by the presence of adsorbate when only the lowest
nanoribbon subband is populated with electrons. That is, the conductance of the
ribbon with the H adsorbate in Fig.\ref{fig:5}(b) near the Dirac point is very
close to $2e^2/h$, the conductance of the pristine ribbon in the same energy
range. This remarkable robustness of transport in the first subband is due to
that subband's unique scattering properties arising from the nature of the
unconfined electron wave function in graphene.\cite{Areshkin07, Yamamoto08} Upon
inclusion of the graphene rehybridization in our model we found in Section
\ref{Dirac} that in addition to the Dirac point scattering resonance for the H
adsorbate shifting to an energy very close to the graphene Dirac point the
resonance also becomes stronger by approximately two orders of magnitude. This
is sufficient to override the relative robustness of transport in the first
graphene nanoribbon subband against scattering so that, unlike in
Fig.\ref{fig:5}(b), the conductance close to the Dirac point (zero energy) in
Fig.\ref{fig:5}(a) for the ribbon with the H adsorbate is very strongly
suppressed relative to that for the ideal ribbon.

\subsubsection{Adsorbed H Atoms and the Density of States}

\begin{figure}[th!]
\includegraphics[keepaspectratio,width=\columnwidth]{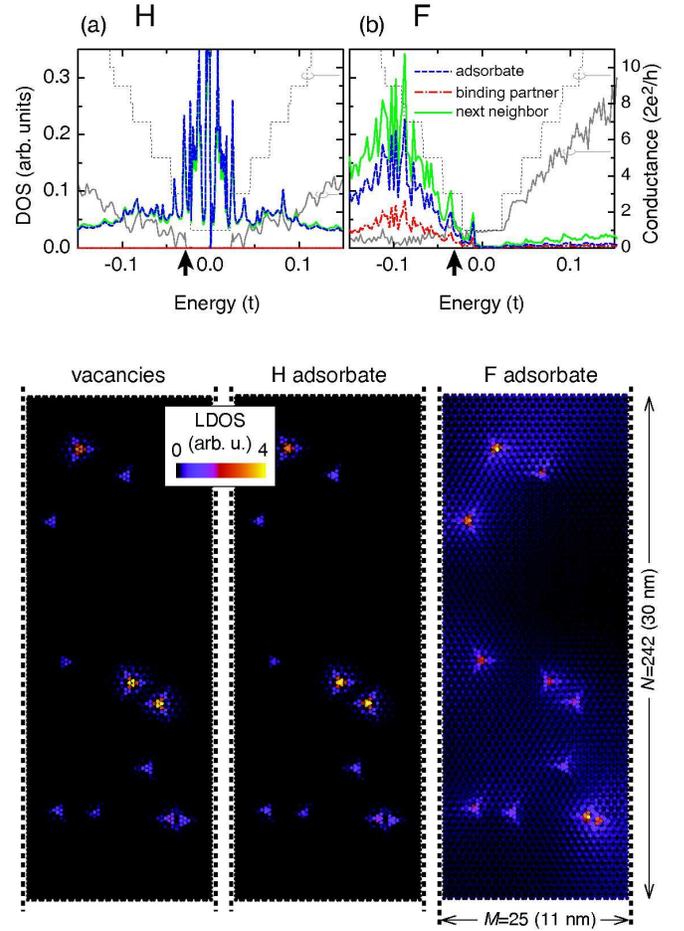}
\caption{(color online) Local density of states (LDOS) averaged over
50 adatoms located at different places in the ribbon vs. electron
energy for a ribbon with adsorbed (a) H and (b) F atoms.
Adsorbate-induced rehybridization of the graphene is included. The
LDOS is shown in red for the $2p_z$ orbital of the C atom to which the
adsorbed atom bonds and in green for the $2p_z$ orbital of a C atom
neighbor of that C atom. The total LDOS on the EMOs associated with
the adsorbed atom (including the contributions of the non-$\pi$
orbitals of the nearby C atoms) is shown in blue. The dotted and solid
grey lines show the conductances of the ideal ribbon and the ribbon
with H or F adatoms, respectively.  Arrows mark the energy chosen for
computation of the LDOS in the bottom panels. The latter are for $15
\times 30$ nm$^2$ fragments of ribbons with interior vacancy defects
(left), H adatoms (middle) and F adatoms (right) located at the same
places. In the middle and right panels the LDOS for the adsorbate EMOs
is combined with the graphene LDOS. For the chosen energy ($-0.03t$)
the missing carbon atoms give rise to very similar electron
localization close to the defects as the presence of the H adatoms
while the LDOS for F is noticeably different.}
\label{fig:7}
 \end{figure}

The presence of the adsorbate leads to electron localization at the adsorbed
atom or molecule as well as at the nearby carbon atoms. Fig. \ref{fig:7} (a)
shows the local density of states (LDOS) for a ribbon with H adatoms averaged
over atoms at different locations. Electron localization is most pronounced for
energies near the Dirac point where the conductance is suppressed most strongly
due to resonant backscattering of electrons. Localization is much stronger at
the H adatom (including its EMOs) than at the $2p_z$ orbital of the carbon atom
to which the H binds. The LDOS for that carbon $2p_z$ orbital  is smaller by
three orders of magnitude than for the H atom. For this reason it is
indistinguishable from the horizontal axis in Fig. \ref{fig:7} (a). The carbon
atoms adjacent to that carbon atom belong to the other graphene sublattice and
exhibit an LDOS as large as that at the H. Strong electron localization is known
to occur in graphene near carbon atom vacancies.\cite{review, Evaldsson08,
Mucciolo09} The lower  left and middle panels in Fig. \ref{fig:7} show the LDOS
for two representative graphene ribbons with interior vacancies and H adatoms,
with these defects located at corresponding sites in the two ribbons. For the
chosen energy $-0.03t$ electrons localize strongly near defects of both types.
The C$_3$ point symmetry of the graphene lattice about the defect sites is
clearly visible in LDOS in the vicinity of each defect. We find the LDOS to
decay according to the power law $1/r$ (not shown in Fig. \ref{fig:7}) for both
defect types although individual defects may exhibit differing LDOS amplitudes.
This power law decay is consistent with the results of other
studies.\cite{Wehling09X, DOSdecay}

\subsection{Ribbons with F Adatoms}
\label{F_Adatoms}

The fluorine adatom bonds covalently to a carbon atom of the graphene lattice
and has two valence orbitals ($2s$ and $2p_z$) that scatter graphene $\pi$
electrons. In the simplest tight-binding model that ignores rehybridization of
the graphene, these two orbitals represent two independent channels for electron
scattering and their contributions to the scattering are additive (for a single
adatom). However, the $2s$ and $2p_z$ orbitals of the F do not contribute
equally: The $2s$ orbital couples more strongly to the carbon $2p_z$ orbital 
since it has the larger value of $| \gamma_{\alpha, C_{2p_z}} |$ as can be seen
in Table \ref{tab:2}. It therefore has the most influence on the electron
transport in the graphene ribbon. The coupling of the $2p_z$ orbital is much
weaker and has a much smaller effect on the conductance; see Fig.
\ref{fig:5}(d).

The Dirac point resonance energy for F on graphene was found in Section
\ref{Dirac} to be  $-0.136t$ in the model that includes rehybridization and
$-0.222t$ in the model that does not. Thus while the effect of rehybridization
of the graphene on the Dirac point resonance of F is important it is not as
drastic as in the case of H. However,  as in the case of adsorbed H, these Dirac
point resonance energies for F agree well with the energies at which the
calculated conductances of ribbons with adsorbed F are most strongly suppressed
in both models as can be seen in Fig. \ref{fig:5}(c) and (d) and Fig.
\ref{fig:6}.

In contrast to the case of the H adsorbate, the conductance is affected only
weakly by the presence of the F adsorbate when only the lowest nanoribbon
subband is populated with electrons for {\em both} models at the adsorbate
concentrations in Fig. \ref{fig:5}(c) and (d).  This is because (unlike for H)
the Dirac point resonances for F are offset significantly in energy from the
Dirac point in {\em both} models.

The calculated LDOS for the F adsorbate for the model that includes the
rehybridization of the graphene is shown in Fig.\ref{fig:7}(b) for the EOMs
asociated with the F adsorbate in blue and for the $2p_z$ orbitals of the C atom
to which the F binds (red) and its C neighbor (green). In each case the LDOS is
strongest around the energy of the F Dirac point resonance. As for H the LDOS
for F is weaker on the $2p_z$ orbitals of the C atom to which the F binds than
either on the neighboring C atoms or on the EMOs of the F adsorbate. The spatial
map of the LDOS for F shown in the lower right panel of Fig.\ref{fig:7} is
qualitatively similar to that for H and for vacancies, but the particular defect
sites showing the strongest LDOS at a given energy are in some cases different
for the F.

Notice also that in Fig.\ref{fig:7}(b) there are peaks in the DOS at positive
energies near subband edge energies that match dips in the conductance that is
also plotted for comparison. These conductance dips are due to enhanced electron
backscattering due to the greater availability of final states for the
scattering process at energies with the higher density of subband states as is
discussed in Ref. \onlinecite{disorder09}.

\subsection{Ribbons with Adsorbed OH Groups}
\label{OH_Ads}

The OH molecule is another monovalent adsorbate with many properties
similar to F adatoms. In the relaxed geometry the O bonds covalently to a C atom
(top site bonding) and the OH chain stands upright relative to the graphene
plane, features found also in Refs. \onlinecite{Robinson08, Wehling09}. As is
seen in Fig. \ref{fig:5}(f), the $2s$ and $2p_z$ O orbitals affect electron
conduction through the graphene ribbons similarly to $2s$ and $2p_z$ orbitals of
F (see Fig. \ref{fig:5}(d)) although for O the orbitals scatter electrons
somewhat more strongly. The H $1s$ orbital in OH molecule has much less
influence because of its large distance from the C and small overlap with the
$2p_z$ C orbital. The total effect of OH adsorbed molecules on the conductance
through the graphene ribbon is qualitatively and quantitatively similar to that
of F; see also Fig. \ref{fig:6}.

The Dirac point resonance energy for OH on graphene was found in Section
\ref{Dirac} to be $-0.089t$ in the model that includes rehybridization and
$-0.194t$ in the model that does not. These numbers again agree reasonably well
with the energies at which the conductances of ribbons with adsorbed OH is
strongly suppressed. The Dirac point resonances for OH occur at energies
intermediate between those for H and F in each model and the same is true of the
energies at which the strongest suppression of the calculated conductance is
seen in Fig. \ref{fig:5} and \ref{fig:6}.

As in the case of the F adsorbate, the conductance is affected only weakly by
the presence of the OH adsorbate when only the lowest nanoribbon subband is
populated with electrons for both models at the adsorbate concentrations in Fig.
\ref{fig:5}(e) and (f), for the same reasons.

\subsection{Ribbons with O Adatoms}
\label{O_Ads}

Oxygen is a bivalent adsorbate that binds simultaneously to two neighboring
carbon atoms (a bridge site). This leads to substantial rehybridization of the
bonding associated with these two carbons that belong to different graphene
sublattices. This, in turn, leads to strong electron scattering and suppression
of electron conduction through the ribbon.

The calculated conductance vs. energy characteristics for ribbons with adsorbed
O are shown in Fig. \ref{fig:5}(g) and  \ref{fig:6} for the model that includes
rehybridization of the graphene and in  Fig. \ref{fig:5}(h) for the model that
does not. The O orbitals affect conduction differently: The O $2s$ orbital
produces strong suppression of the conductance at positive energies, but the
$2p_z$ orbital suppresses the conductance much less and this occurs mainly at
negative energies. The effect of the O $2p_x$ orbital is weaker still. The
resulting low conductance at positive energies is a consequence in part of the O
adsorbate binding to two carbon atoms belonging to two graphene sublattices
unlike the F and OH that bind to a single C atom and exhibit low conductance at
negative energies. We note however that in general the sign of the energy at
which low conductance due to adsorbate scattering occurs depends not only on the
number of C atoms to which the adsorbate binds but also on the values of the
model tight binding parameters $\epsilon_\alpha$ and $\gamma_{\alpha j}$. For
example, if the absolute values of $\gamma_{\alpha j}$ were substantially
smaller than our estimate the low conductance region for O would be at {\em
negative} energies relative to the Dirac point of the ribbon.

As for the other adsorbates discussed above both the sign and magnitude of the
energy at which the conductance of the ribbon with the O adsorbate is suppressed
agree well with the energies at which the Dirac point resonances for the O
adsorbate were found to occur in Section \ref{Dirac}, i.e., $0.112t$ and
$0.090t$ in the model that includes the graphene rehybridization and $0.046t$
for that which does not.

Another striking feature of Fig.  \ref{fig:6} is the prominent conductance
maximum for the ribbon with the O adsorbate at energies near $ 0.55t$. The
energy at which the conductance maximum  occurs coincides with the energy of the
antiresonance (i.e., deep minimum) of the norm of the $T$-matrix (that describes
scattering of graphene electrons due to an adsorbed O atom) that was found in
Section \ref{Dirac} in the model that includes the O adsorbate-induced
rehybridization of the graphene.

Since weak electron scattering is normally associated with high conductance this
match between the conductance maximum and the $T$-matrix antiresonance is
intuitively reasonable. It demonstrates once again the close relationship
between the $T$-matrix theory of Sections \ref{Tmatrix} and \ref{Dirac} and
transport in graphene ribbons with chemisorbed species that has already been
illustrated by the agreement found above between the electron energies at which
$T$-matrix resonances and strong suppression of the ribbon conductances occur.

\subsection{Conductance Asymmetry Relative to the Dirac Point of Graphene}
\label{asym}

In Section \ref{Dirac} we found the Dirac point resonances for F, OH and O to be
offset in energy from the Dirac point of graphene and in Sections
\ref{F_Adatoms}, \ref{OH_Ads} and \ref{O_Ads} we showed the electron Fermi
energies at which the lowest conductances of ribbons with these adsorbates occur
to be offset from the Dirac point of graphene similarly.

Such asymmetric conduction relative to the Dirac point has been discussed in
Ref. \onlinecite{Robinson08} for the case of infinite two-dimensional graphene.
Suppression of the conductivity on one side of the Dirac point with only a weak
effect on conduction on the other was explained by a local energy-dependent
scattering potential due to the adsorbate.\cite{Robinson08} Over a range of
energies the conductivity was found to be small while rising linearly outside
that range. Qualitatively similar results for 2D graphene were subsequently 
reported by others.\cite{Wehling10, Skrypnyk10} A transport gap around the 
impurity resonance energy had also 
been predicted earlier for a generic model of infinite 2D graphene with 
point defects\cite{Skrypnyk06} although this prediction was not 
supported by transport
calculations.\cite{Skrypnyk06}

This behavior resembles our results for the conductance in the graphene ribbons
with F, OH and O adsorbates but with a qualitative difference: As will be
discussed in Section \ref{Quantization}, at moderately low temperatures the
graphene ribbons exhibit quantized conductance steps superposed on the otherwise
linearly rising conductance due to enhanced electron backscattering at the edges
of the subbands of the ribbon.

\subsection{Conductance Quantization}
\label{Quantization}

\begin{figure}[t]
\includegraphics[keepaspectratio,width=0.8\columnwidth]{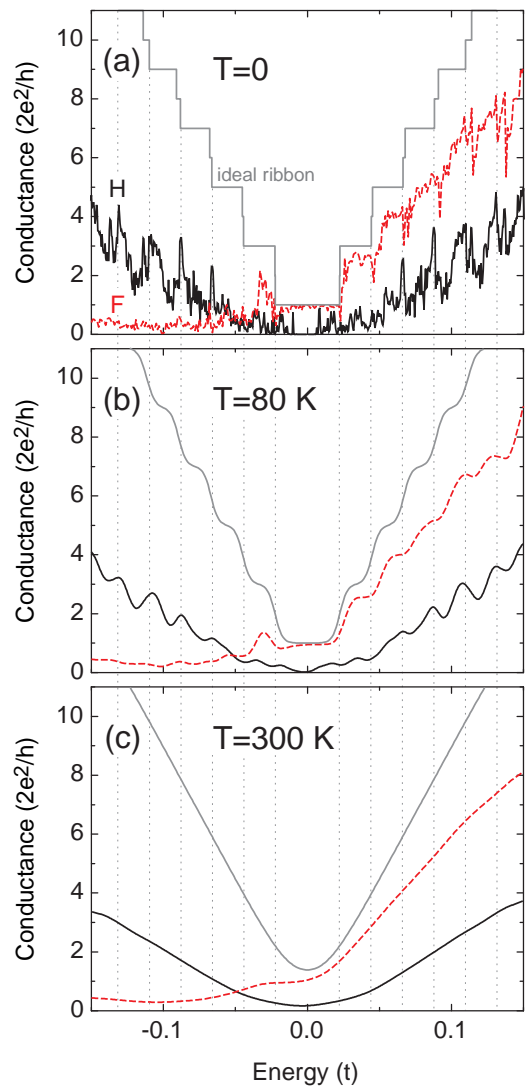}
\caption{(Color online) The calculated conductances of ideal graphene ribbons
with no adsorbate (solid grey) with H adatoms (solid black) and with F adatoms
(dashed red). The temperature is 0, 80 and 300 K in (a), (b) and (c)
respectively. Adatom concentrations are $p=10^{-4}$ as in Fig.\ref{fig:5}.
Adsorbate induced graphene rehybridization is included in the model. Ribbon
width $W=30$ nm; length $L=500$ nm.  $t=2.7$ eV.}
\label{fig:8}
\end{figure}

In the experimental study of Lin et al.\cite{Lin08} conductance quantization in
the form of conductance steps of equal height was observed in graphene
nanoribbon samples with conductances much smaller than $2e^2/h$ in a range of
moderately low temperatures as a gate voltage applied to the sample was varied.
In a previous paper\cite{disorder09} we showed theoretically that conductance
quantization of this kind should occur in graphene nanoribbons with interior
carbon atom vacancies even if comparable amounts of other defects such as edge
disorder and long range potentials due to charged defects are also present.
However, because of the sample preparation techniques used by Lin et
al.\cite{Lin08} adsorbed H may well have been present in their samples.

As we have already noted in Section  \ref{H_conductance} (and is demonstrated
very clearly in Fig \ref{fig:5} (a)) the calculated zero temperature conductance
of graphene ribbons with adsorbed H atoms is very similar both qualitatively and
quantitatively to that of the ribbon with the same concentration of carbon atom
vacancies. In particular, the properties of the conductance characteristics of
ribbons with vacancies that have been shown\cite{disorder09} to give rise to
conductance quantization of the kind observed by Lin et al.\cite{Lin08} are also
exhibited by the conductance characteristic of the ribbon with H adatoms in Fig
\ref{fig:5} (a). These properties are

(i) Pronounced sample-specific conductance fluctuations that are manifestation
of quantum interference.\cite{disorder09}

(ii) If the conductance fluctuations are ignored, the conductance is seen to
scale down uniformly overall due to scattering by the adsorbate, i.e., in a
similar way for all subbands.

(iii) The conductance shows a pronounced dip whenever a new subband becomes
available for electron propagation. As is discussed in Ref.
\onlinecite{disorder09}, this is because of enhanced electron backscattering by
the defects at subband edges.

Because of these three properties we expect graphene nanoribbons with adsorbed H
to exhibit equally spaced conductance steps similar to those observed
experimentally by Lin et al,\cite{Lin08} for the same reasons and under the
similar conditions (discussed in Ref. \onlinecite{disorder09}) as do graphene
nanoribbons with carbon atom vacancies. That is, conductance steps of equal
height should be observed even in samples with sufficiently high adsorbate
concentrations for the ribbons to have conductances much smaller than
$2e^{2}/h$, the conductance steps should break up into random conductance
fluctuations as the temperature approaches zero Kelvin, and the conductance
steps should become completely smeared out by thermal broadening at temperatures
substantially larger than the subband spacing of the ribbons. This is indeed
seen in Fig. \ref{fig:8} where we show the calculated conductance of a ribbon
with H adatoms at different temperatures: Regular conductance features are seen
at 80K in Fig. \ref{fig:8} (b). They break up into universal conductance
fluctuations at 0K in Fig. \ref{fig:8} (a) and are completely smeared out
thermally at 300K in Fig. \ref{fig:8} (c).

Thus electron scattering due to a low concentration of H atoms adsorbed on the
graphene nanoribbons in our model that takes into account the adsorbate-induced
rehybridization of the graphene provides an alternative and equally satisfactory
explanation of the conductance quantization observed by Lin et al. \cite{Lin08}
to that \cite{disorder09} provided by electron scattering by carbon atom
vacancies in the interior of the ribbon.

It is also possible that adsorbed F, OH and O were present in the samples of Lin
et al.\cite{Lin08} The three properties (i), (ii) and (iii) are also shared by
ribbons with F, OH and O adsorbates for positive, positive and negative energies
respectively, as can be seen in Fig. \ref{fig:5} (c), (e) and (g). (Notice also
the density of states maxima at subband edges that are
responsible\cite{disorder09} for enhanced electron backscattering and hence for
a conductance dip whenever the electron Fermi level crosses a subband edge
(i.e., property (iii) above) that are clearly visible at positive energies in
Fig. \ref{fig:7}(b)).  Therefore, conductance quantization of the kind observed
by Lin et al. \cite{Lin08} should also occur under appropriate conditions for
{\em some} ranges of the gate voltage in graphene ribbons with adsorbed F, OH
and O. Theoretical results for H and F adatoms at various temperatures are shown
in Fig. \ref{fig:8}. Note that the nearly perfect transmission of electrons in
the first subband through the ribbons with F, OH and O seen in Fig. \ref{fig:5}
and \ref{fig:7} is specific to armchair ribbons of particular
widths\cite{Dresselhaus96} that are metallic and does not occur for insulating
ribbons. However, unlike for adsorbed H, for adsorbed F, OH and O, the Dirac
point has a distinctive signature in the calculated conductance that does not
coincide with the conductance minimum that is due to the scattering resonance(s)
associated with the adsorbate. Such a signature is not evident in the
experimental data of Lin et al.\cite{Lin08}.

In the experimental data of Lin et al.\cite{Lin08} there is an offset of about 3
Volts between the gate voltage at which the conductance minimum occurs and zero
gate voltage. The offsets between the energies at which the conductance minima
occur in our calculations and the Dirac point energy for F, OH and O are all
smaller than 0.3 eV, the largest being for the F adsorbate. However, we expect
other mechanisms such as charged impurities in the insulating spacer between the
ribbon and gate electrode, contact potentials between the spacer and gate and/or
between the spacer and ribbon to contribute significantly to the the
experimentally observed gate voltage at which the conductance minimum occurs.

\subsection{Adsorbate Induced Renormalization of the Dirac Point and 
Subband Edge Energies}
\label{renorm}  

We note that while for F, OH and O the conductance minimum is displaced in
energy from the Dirac point of the pristine ribbon, this should {\em not} be
interpreted as a shift of the Dirac point of the ribbon due to interaction with
the adsorbate;\cite{SkrypnykDshift} it arises almost entirely from 
suppression of the conductance by
enhanced electron scattering near the conductance minimum.

The actual shift of the Dirac point energy 
(and subband edge energies) of the
ribbon due to interaction of the ribbon with the adsorbate is much smaller and
can be estimated perturbatively as follows: If we consider the coupling term
$H_c = \sum_{\alpha, j} \gamma_{\alpha j} \left( d_{\alpha}^{\dag }a_{j}+h.c.
\right)$ between the adsorbate and the $\pi$ states the ribbon in the tight
binding Hamiltonian Eq. (\ref{eq:tbhamiltonian}) as a perturbation then in
second order perturbation theory this coupling implies a shift $\Delta_{ks}$ in
the energy of the subband state $|ks\rangle$ given by
\begin{equation}
	\Delta_{ks} = \sum_{\alpha, j} |\langle\psi_{\alpha j}|H_c |ks\rangle|^2 
	/(\epsilon_{ks}-\epsilon_\alpha)
\label{eq:energyshift}
\end{equation}
 where $\epsilon_{ks}$ is the unperturbed energy of state $|ks\rangle$. For
 states $|ks\rangle$ that are much closer in energy to the Dirac point of the
 ribbon than are the orbital energies $\epsilon_\alpha$ of the adsorbate we can
 approximate $\epsilon_{ks}-\epsilon_\alpha \sim -\epsilon_\alpha$ in the
 denominator of Eq. \ref{eq:energyshift}. For OH and F adsorbates  that bind to
 the graphene over a single C atom and assuming that they are randomly
 distributed over the ribbon we can on average approximate $|\langle\psi_{\alpha
 j}|H_c |ks\rangle|^2$ by $|\gamma_{\alpha j}|^2 /N$ where $N$ is the number of
 carbon atoms in the ribbon. With these approximations Eq. \ref{eq:energyshift}
 becomes

\begin{equation}
	\Delta_{ks} \sim -p  \sum_{\alpha} |\gamma_{\alpha j}|^2 /\epsilon_\alpha
	\label{eq:approxshift}
\end{equation}
where $p$ is the concentration of the adsorbate and the sum is over the extended
molecular orbitals of a single adsorbed atom or molecule.  Inserting $p=10^{-4}$
and the values of $\gamma_{\alpha j}$ and $\epsilon_\alpha$ from Table
\ref{tab:1} in Eq. \ref{eq:approxshift} we find $\Delta_{ks} \sim  4.4 \times
10^{-4} t$ and $ 6.3 \times 10^{-4} t$ for F and OH respectively. These shifts
are very small justifying our use of perturbation theory and demonstrating that
the shift of the Dirac point energy due to this mechanism is very small for the
adsorbate concentrations considered in this article.

\begin{figure}[t]
\includegraphics[keepaspectratio,width=\columnwidth]{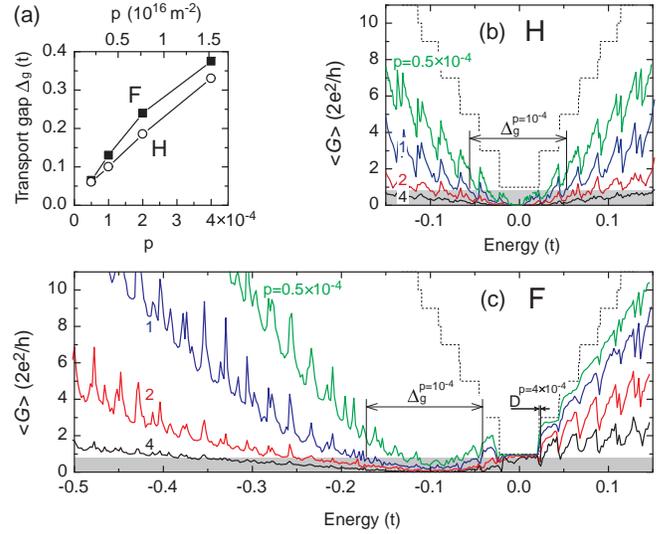}
\caption{The averaged conductances of graphene ribbons with different
concentrations $p$ of adsorbed atoms of (b) H and (c) F averaged over different
locations of the adsorbed atoms. The results for $p = 0.5 \times 10^{-4}, 1
\times 10^{-4}, 2 \times 10^{-4}$ and $ 4 \times 10^{-4} $ are shown in green,
blue, red and black, respectively. The averaging in each plot is over 10
arrangements of the positions of the adsorbed atoms. The transport gap
$\Delta_g$ is estimated as the energy interval where $G<0.9 \times 2e^2/h$. (a)
$\Delta_g$ increases approximately linearly with $p$ at low $p$. $D \sim 1.5
\times 10^{-3} t$ in (c) is the offset between the second subband edge for the
ideal ribbon with no adsorbate and the center of the conductance dip due to
enhanced electron backscattering when the (renormalized) edge of the second
ribbon subband crosses the electron Fermi level in the ribbon with a
concentration $p = 4 \times 10^{-4}$ of adsorbed F atoms; see the last paragraph
of Section \ref{renorm}. Note that $p=10^{-4}$ corresponds to $3.8\times
10^{15}$ adsorbed atoms per square meter. $t=2.7$ eV.}
\label{fig:9}
\end{figure}

Despite its small size the energy shift $\Delta_{ks}$ has a clear signature in
the results of our transport calculations: As we have discussed in Ref.
\onlinecite{disorder09} and in the preceding subsections of Section
\ref{results} of the present article, the conductance of a ribbon with disorder
(including that due to randomly located adsorbed moieties) shows a pronounced
dip whenever a new subband becomes available for electron propagation due to
enhanced electron backscattering by the defects at subband edge energies. Given
that the interaction between the ribbon and adsorbate renormalizes the subband
edge energies by the amount $\Delta_{ks}$ it is to be expected that the energies
at which the conductance dips due to scattering by an adsorbate occur should be
offset from the subband edge energies of the {\em pristine} ribbon also by
approximately $\Delta_{ks}$. This is indeed what we find. For example, for F
adatoms on a ribbon at a concentration $p=4 \times 10^{-4}$ Eq.
\ref{eq:approxshift} yields $\Delta_{ks} = \sim 1.8  \times 10^{-3}t$ which
agrees reasonably well with the displacement $D \sim 1.5  \times 10^{-3}t$ of
the center of the conductance dip from the second subband edge of the pristine
graphene ribbon that is seen for the $p=4 \times 10^{-4}$ F adsorbate in Fig.
\ref{fig:9}(c).

\subsection{The Transport Gaps and Their Dependence on Adatom Concentration}
\label{gapsvsconc}

Figure \ref{fig:9} shows the conductances of graphene ribbons with different
concentration of H and F adatoms. As the adsorbate concentration increases the
conductance decreases and a wider transport gap in which the ribbon is
effectively an insulator opens centered near the Dirac point for H and at
negative energies for adsorbed F. The width $\Delta_g$ of the transport gap
(that we define arbitrarily as the energy range where the ribbon conductance is
less than $0.9 \times 2e^2/h$) grows linearly with the adatom concentration at
low concentrations. It worth noting that $\Delta_g$ does not depend on the
ribbon width, which rules out adsorbate impurities as the principal source of
the transport gap in the experiment in Ref. \onlinecite{Han07}.

Transport gaps gaps have been discussed previously in Ref.
\onlinecite{Pereira2008} for  graphene with uncompensated vacancies. A $
\Delta_g \sim \sqrt{p}$ dependence was found over a wide range of defect
concentrations $0<p<0.2$. Note, however, that for vacancies randomly distributed
over the two sublattices no transport gap was found in Ref.
\onlinecite{Pereira2008}. Transport gaps have also been predicted for graphene
ribbons substitutionally doped with boron,\cite{Biel09} however, whether the
predicted transport gaps are related to Dirac point 
resonances was not discussed.\cite{Biel09}

\section{Discussion}
\label{Discussion}

In this article we have formulated a tight binding theory of the Dirac point
resonances due to adsorbed atoms and molecules on graphene based on the standard
tight binding model of the graphene $\pi$-band electronic structure and the
extended H\"{u}ckel model of the adsorbate and the adsorbate-induced local
$sp^3$ rehybridization of the graphene. We generalized previous theories of the
effective Hamiltonians of graphene with impurities to the case of adsorbate
species with multiple extended molecular orbitals and bonding to more than one
graphene carbon atom, and obtained accurate analytic expressions for the Green's
function matrix elements that enter the $T$-matrix theory of Dirac point
resonances. This generalization makes the extended H\"{u}ckel model (and
potentially other tight binding models as well) into a powerful tool for
studying the Dirac point resonances induced by many different adsorbates on
graphene.  Furthermore this theory makes it practical to carry out sophisticated
electronic quantum transport calculations for graphene nanoribbons tens of
nanometers wide (such as are being realized  in present day experimental
studies) with adsorbates covalently bonded to the ribbon. We applied the above
theoretical approach to H, F, OH and O adsorbates on graphene whose relaxed
geometries we calculated with {\em ab initio} density functional theory. For
each of these adsorbates we found a strong scattering resonance near the Dirac
point of graphene, the strongest by far being for the H adsorbate. Treating the
valence orbitals of the adsorbed species and the 2$s$, 2$p_x$ and 2$p_y$ valence
orbitals of the nearby carbon atoms theoretically in a unified way was necessary
in order to obtain reliable results. We also extracted from these calculations a
minimal set of tight binding parameters that make it possible to efficiently
model adsorbate-induced electron scattering and its effect on electron transport
in graphene and graphene nanostructures.

In particular, the minimal tight binding models that we developed make it
possible to model the effect of adsorbates on transport in graphene nanoribbons
tens of nanometers wide and hundreds of nanometers long that are at present
being realized experimentally. We have presented realistic electronic quantum
transport calculations for such nanoribbons with adsorbed H, F, OH and O. As
well as the carbon $\pi$ band electronic structure of the graphene nanoribbons
our theory includes the effects of the local partial rehybridization of the
graphene ribbon from the $sp^2$ to $sp^3$ electronic structure that occurs when
H, F, OH or O bonds covalently to the ribbon. This is necessary in order for the
model to describe correctly the scattering resonances that are induced in the
graphene ribbons near the Dirac point by the presence of these adsorbates. We
find that these Dirac point resonances play a dominant role in quantum transport
in ribbons with these adsorbates: Even at low adsorbate concentrations of
$10^{-4}$ adsorbed atoms or molecules per carbon atom, in the ribbons that we
study the conductance of the ribbon is strongly suppressed and a transport gap
is formed for electron Fermi energies in the vicinity of the energy of the
resonance. For the H adsorbate this transport gap is centered very close to the
Dirac point energy of the ribbon (as it is for ribbons with interior carbon atom
vacancies) while for F and OH it is centered below the Dirac point and for O it
is centered above the Dirac point. These predictions can be tested
experimentally by tuning the position of the Fermi level in the ribbon relative
to its Dirac point by means of a variable applied gate voltage. For each of
these adsorbed species we find a pronounced dip in the low temperature
conductance of the ribbon when the electron Fermi level crosses the edge of an
electronic subband of the ribbon due to enhanced electron backscattering, and
the conductance to be suppressed equally on average in every subband. This
implies that graphene nanoribbons with H, F, OH and O adsorbates and
conductances even a few orders of magnitude smaller than $2e^2/h$ should (for
appropriate ranges of a back gate voltage) exhibit equally spaced conductance
steps at moderately low tempertaures similar to those that have been observed by
Lin et al.\cite{Lin08} experimentally and that have recently been predicted
theoretically\cite{disorder09} for graphene nanoribbons with interior carbon
atom vacancies.

Experiments testing these predictions by observing the Dirac point resonances in
lateral transport through well characterized graphene nanoribbons with H, F, OH
and O adsorbates intentionally deposited at known concentrations would be of
interest. Vertical transport measurements directed at detecting the Dirac point
resonances and measuring the energies at which they occur more directly with the
help of scanning tunneling spectroscopy of atoms and molecules adsorbed on
graphene and graphene ribbons would also be interesting, especially in view of
the often conflicting predictions of the values of some of these energies that
have been made by various theory groups.

\begin{acknowledgments}
We thank A. Saffarzadeh for his helpful comments. This work was supported by
NSERC, CIFAR and WestGrid. 
\end{acknowledgments}


\begin{thebibliography}{99}
\bibitem{no_overlap} If the calculation for the model in Fig. \ref{noCsigma} is
repeated but neglecting the overlaps $ \sigma_{\alpha j}$ between the EMOs and
graphene $2p_z$ orbitals we find ${\epsilon_{\mathrm{DR}}}=  -0.220t,  -0.189t, 
-0.131t$ and $   0.045t$ for F, OH, H and O, respectively, values that differ by
a few percent from those obtained in the corresponding calculations that include
the overlaps $ \sigma_{\alpha j}$. These differences and the corresponding
differences in the strengths of the resonances are quite small because both the
Dirac point resonance energies ${\epsilon_{\mathrm{DR}}}$ and the overlaps $
\sigma_{\alpha j}$ are small and therefore the energy-dependent correction terms
$-\epsilon \sigma_{\alpha j}$ to Hamiltonian matrix elements $ \gamma_{\alpha
j}$ that are due to the overlaps (discussed at the end of Section \ref{tbH}) are
small at the resonance energies.
\bibitem{scaling}The extended H\"{u}ckel model as parameterized in Ref.
 \onlinecite{yaehmop} (after the application of Eq. (\ref{eq:shift})) yields the
 value 2.02 eV for $t_{ij}$ for nearest neighbor carbon atoms of undistorted
 pristine graphene. For consistency with the standard tight binding
 model\cite{review} of pristine graphene in which the nearest neighbor hopping
 parameter $t_{ij}=t=2.7$ eV we therefore scale the values of  $t_{ij}$ obtained
 from the extended H\"{u}ckel model for nearest neighbor carbon atoms in the
 graphene structure that has been distorted by the presence of the adsorbate by
 a factor of 2.7/2.02. No scaling is applied to the values of any other tight
 binding parameters that we obtain from extended H\"{u}ckel theory.
\bibitem{SkrypnykDshift}The Dirac point shift has been considered previously in Ref. \onlinecite{Skrypnyk06} for a simple generic model of point defects in infinite 2D graphene.
\end{thebibliography}
\end{document}